\documentclass[twocolumn,prl]{revtex4}%
\usepackage[dvipdfmx]{graphicx}%
\usepackage{amsmath,amssymb}%
\usepackage{amsfonts}
\usepackage{amssymb}
\usepackage{mathrsfs}
\usepackage{color}
\usepackage{bm}

\def\s{{\sigma}}
\def\e{{\epsilon}}
\def\k{{ {\bm k} }}
\def\p{{ {\bm p} }}
\def\q{{ {\bm q} }}

\def\0{{ {\bm 0} }}
\def\w{{\omega}}
\def\a{{\alpha}}

\def\r{{ {\bm r} }}

\allowdisplaybreaks[4]

\begin{document}
\title{
Mechanism of exotic density-wave 
and beyond-Migdal unconventional superconductivity 
in kagome metal AV$_3$Sb$_5$ (A=K, Rb, Cs)
}
\author{
Rina Tazai$^*$, Youichi Yamakawa, Seiichiro Onari, and Hiroshi Kontani$^*$
}
\date{\today }

\begin{abstract}
{\bf
Exotic quantum phase transitions in metals, 
such as the electronic nematic state,
have been discovered one after another and found to be universal now.
The emergence of unconventional density wave (DW) order in frustrated 
kagome metal AV$_3$Sb$_5$ and its interplay with 
exotic superconductivity attract increasing attention.
We find that the DW in kagome metal is the bond-order,
because the sizable inter-site attraction is caused by 
the quantum interference among paramagnons.
This mechanism is significant in kagome metals 
since the geometrical frustration prohibits the freezing of paramagnons.
In addition, we uncover that moderate bond-order fluctuations
mediate sizable pairing glue, and this mechanism gives rise to 
both singlet $s$-wave and triplet $p$-wave superconductivity.
Furthermore,
characteristic pressure-induced phase transitions in CsV$_3$Cb$_5$
are naturally understood by the present theory.
Thus, both the exotic density wave and the superconductivity 
in geometrically frustrated kagome metals
are uniquely explained by the quantum interference mechanism.
}

\vspace{10mm}
$^*$Corresponding author. Email: tazai@s.phys.nagoya-u.ac.jp (R.T.); kon@slab.phys.nagoya-u.ac.jp (H.K.)

\vspace{10mm}
{\bf Teaser:}
The exotic density wave, superconductivity and their interplay in novel kagome metals are revealed based on the quantum interference mechanism.

\end{abstract}

\address{
Department of Physics, Nagoya University,
Furo-cho, Nagoya 464-8602, Japan. 
}

\sloppy

\maketitle

\section{Introduction}
\label{sec:Intro}

Rich quantum phase transitions 
in strongly correlated metals with 
multiple degrees of freedom and geometrical frustration
have been discovered one after another recently
\cite{Fradkin-rev2012,Davis-rev2013,Shibauchi-rev2020,Tazai-rev2021}.
To understand such rich phase transitions,
a significant ingredient is various
quantum interference processes among different fluctuations 
\cite{Kontani-PRB2011,Onari-SCVC,Tsuchiizu1,Yamakawa-Cu,Yamakawa-FeSe,Onari-FeSe,Chubukov-PRX2016,Fernandes-rev2018}.
The recent discovery of unconventional density-wave (DW) order 
and exotic superconductivity in kagome lattice metal AV$_3$Sb$_5$  (A=K, Rb, Cs)
have triggered enormous number of experimental
 \cite{kagome-exp1,kagome-exp2,kagome-P-Tc1,kagome-P-Tc2,kagome-P-Tc3,kagome-nodal1,kagome-nodal2,kagome-full,kagome-full2,first-CDW}
and theoretical
\cite{Balents2021,first-Tc,Thomale2021,Neupert2021,QSi2021,Lin2021}
researches.
Especially, the nontrivial interplay
between density-wave and superconductivity
in highly frustrated kagome metals
has attracted increasing attention in condensed matter physics.


At ambient pressure,
AV$_3$Sb$_5$ exhibits charge-channel DW order 
at $T_{\rm DW}=78$, $94$ and $102$ K for A=K, Cs and Rb, respectively
\cite{kagome-exp1,kagome-exp2,NMR2,NMR1}.
Below $T_{\rm DW}$, $2\times2$ (inverse) star of David pattern 
was observed by STM studies
\cite{STM1,STM2}.
The absence of acoustic phonon anomaly at $T_{\rm DW}$
\cite{Kohn}
would exclude DW states due to strong electron-phonon coupling.
As possible electron-correlation-driven DW orders,
charge/bond and loop-current (LC) orders 
\cite{Balents2021,Thomale2021,Lin2021,Thomale2013,SMFRG,kagome-MF}
have been proposed theoretically,
mainly based on the extended Hubbard model
with the on-site ($U$) and the nearest-neighbor site ($V$) Coulomb interactions. 
However, when $V\ll U$ due to Thomas-Fermi screening,
previous studies predicted strong magnetic instability,
in contrast to the small spin fluctuations in AV$_3$Sb$_5$ at $T_{\rm DW}$
\cite{NMR1,NMR2,mSR}.

Below $T_{\rm DW}$, exotic superconductivity 
occurs at $T_{\rm c}=1 \sim 3$ K at ambient pressure
\cite{kagome-nodal1,kagome-nodal2}.
The finite Hebel-Slichter peak in $1/T_1T$ \cite{NMR2}
and the absence of the impurity bound-state below $T_{\rm c}$ 
\cite{impurity}
indicate the singlet $s$-wave superconducting (SC) state.
On the other hand,
the possibilities of triplet pairing state 
\cite{triplet} 
and nematic SC state
\cite{nematic-SC1,nematic-SC2}
have been reported.
In addition, topological states
have been discussed intensively
\cite{topo-SC}.
Under pressure, $T_{\rm DW}$ decreases and 
vanishes at the DW quantum critical point (DW-QCP) 
at $P\sim2$GPa.
For A=Cs, $T_{\rm c}$ exhibits a nontrivial 
double SC dome in the DW phase, 
and the highest $T_{\rm c}\ (\lesssim10 {\rm K})$ 
is realized at the DW-QCP
\cite{kagome-P-Tc1}.
In addition, theoretical phonon-mediated $s$-wave $T_{\rm c}$
is too low to explain experiments
\cite{first-Tc}.
Thus, non-phonon SC state due to DW fluctuations
\cite{Tazai-HF-SC1}
is naturally expected in AV$_3$Sb$_5$.

\begin{figure}[htb]
\includegraphics[width=.87\linewidth]{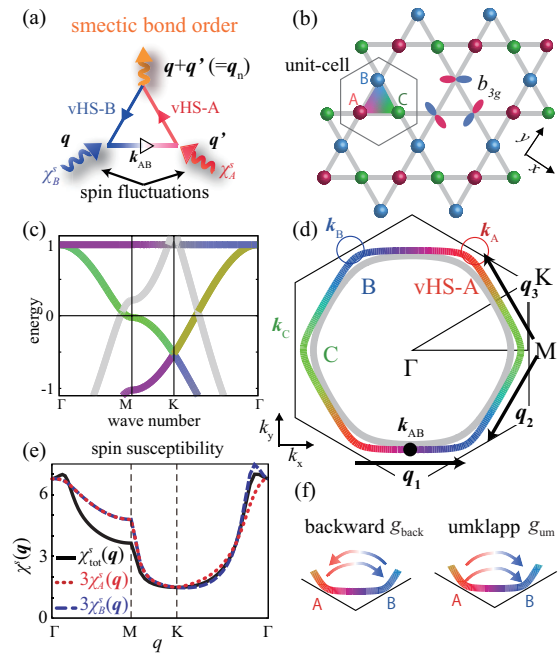}
\caption{
{\bf Interference mechanism, FS and three vHS points in kagome model:}
(a) Smectic order at wavevector $\q_n=\q+\q'$
driven by the paramagnon interference mechanism.
(b) Kagome lattice structure of the vanadium-plane.
Three $b_{3g}$-orbitals $A,B,C$ (and three $b_{2g}$-orbitals $A',B',C'$)
are located at A, B, C sites, respectively. 
(c) Band structure and (d) FSs at $n=3.8$.
The outer (inner) FS are composed of $b_{3g}$- ($b_{2g}$-) orbitals.
Three vHS points $\k_{\rm A}$, $\k_{\rm B}$ 
and $\k_{\rm C}$ are respectively composed of 
$A$ (red), $B$ (blue), and $C$ (green) orbitals.
The inter-vHS nesting vectors $\q_n$ ($n=1,2,3$) are shown.
All $b_{2g}$-orbitals are expressed as gray color.
(e) $\chi^s_A(\q)$, $\chi^s_B(\q)$, and
$\chi^s_{\rm tot}(\q)$ in the RPA.
(f) Backward and umklapp scattering between different vHS points.
These processes are caused by paramagnon interference mechanism.
(See Fig. \ref{fig:fig3} for detail.)
}
\label{fig:fig1}
\end{figure}

The current central issues would be summarized as:
(i) Origin of the DW state and its driving mechanism,
(ii) Mechanism of non-phonon SC state, and
(iii) Interplay between DW and superconductivity.
Such nontrivial phase transitions would be naturally explained 
in terms of the quantum interference mechanism.
The interference among spin fluctuations 
\cite{Kontani-PRB2011,Onari-SCVC,Onari-FeSe,Yamakawa-FeSe,Onari-AFN,Chubukov-W}
(at wavevectors $\q$ and $\q'$)
give rise to unconventional DW at $\q+\q'$,
which is shown in Fig. \ref{fig:fig1} (a).
This mechanism has been applied to explain 
the orbital/bond-orders in various metals
\cite{Tsuchiizu4,Kawaguchi,Tazai-rev2021,Hirata,Tazai-CeB6}.
It is meaningful to investigate 
the role of the paramagnon interference in kagome metals
because the geometrical frustration
prohibits the freezing of paramagnons.
Its lattice structure, band dispersion, and Fermi surface (FS) 
with three van Hove singularity (vHS) points 
are shown in Figs. \ref{fig:fig1} (b), (c) and (d), respectively.

In this paper, we present a unified explanation for 
the DW order and exotic SC state in geometrically frustrated
kagome metal AV$_3$Sb$_5$
that is away from the magnetic criticality,
by focusing on the beyond-mean-field electron correlations.
The DW is identified as the ``inter-sublattice bond-order'' 
that preserves the time-reversal-symmetry.
It originates from the paramagnon interference mechanism
that provides sizable inter-sublattice backward and umklapp scattering.
In addition, we uncover that the smectic DW fluctuations
induce sizable ``beyond-Migdal'' pairing interaction
that leads to the singlet $s$-wave SC state.
The triplet $p$-wave state also appears
when spin and DW fluctuations coexist.
The origins of the star of David order, the exotic superconductivity,
and the strong interplay among them are uniquely explained
based on the quantum interference mechanism.
This mechanism has been overlooked previously.

In the discussion section, we study the
$P$-$T$ phase diagram and the impurity effect on the SC states.
The commensurate-incommensurate (C-IC) DW transition 
is obtained at $1$GPa based on the realistic model constructed by the 
first-principles study.
Based on this C-IC transition scenario,
we put the following theoretical predictions (i)-(iv):
(i) For $ 0 \le P<1$GPa, the commensurate DW emerges.
(ii) For $ P>1$GPa , the DW state turns to be incommensurate due to 
the pressure-induced self-doping on the $b_{3g}$-orbital FS 
(about 1.5 \%).
(iii) As the highest-$T_{\rm c}$ dome at $P\sim 2$GPa,
anisotropic $s$-wave SC state is realized by
the bond-order fluctuations.
(iv) In another SC dome at $ P\sim 0.7$GPa,
both $s$- and $p$-wave states can emerge since the
spin and bond-order fluctuations would be comparable. 
Thus, impurity-induced $p$-wave to $s$-wave transition may occur.
The present key findings 
will stimulate future experiments on AV$_3$Sb$_5$.
 






\section{Results}
\label{sec:Results}

\subsection{Bandstructure with three vHS points}

We analyze the following six orbital kagome lattice Hubbard model 
introduced in Ref. \cite{Thomale2021}.
It is composed of three $b_{3g}$-orbitals ($A,B,C$)
and three $b_{2g}$-orbitals ($A',B',C'$).
Orbitals $A$ and $A'$ are located at A site, for instance.
The kinetic term in $\k$-space is given as
\begin{eqnarray}
H_0=\sum_{\k,l,m,\s} \epsilon_{lm}(\k) c^{\dagger}_{\k,l,\sigma} c_{\k,m,\sigma} ,
\end{eqnarray}
where $l,m=A,B,C,A',B',C'$.
In the paper the unit of energy (in Coulomb interaction,
hopping integral, and temperature) is eV.
The nearest-neighbor hopping integrals are
$t_{b{3g}}=0.5$, $t_{b{2g}}=1$ and $t_{b{3g},b{2g}}=0.002$,
and the on-site energies are $E_{b{3g}}=-0.055$ and $E_{b{2g}}=2.17$
\cite{Thomale2021}.
In the numerical study, it is convenient to 
analyze the six-orbital triangular lattice model 
in Fig. \ref{fig:figS1} in the Supplemental Materials (SM) A,
which is completely equivalent to 
the kagome metal in Fig. \ref{fig:fig1} (b).
In the $b_{3g}$-orbital band shown in Fig. \ref{fig:fig1} (d),
each vHS point (A, B and C) is
composed of pure orbital ($A$, $B$ and $C$), 
while the point $\k_{\rm AB}=(\k_{\rm A}+\k_{\rm B})/2$  
is composed of orbitals $A$ and $B$.
The present $b_{3g}$-orbital FS in the vicinity of three vHS points,
on which the pseudogap opens below $T_{\rm DW}$
\cite{ARPES-CDWgap,ARPES-Lifshitz,ARPES-CDWgap2},
well captures the observed FS
\cite{STM1,ARPES-VHS,ARPES-band}.

Next, we introduce the ``on-site Coulomb interaction'' term $H_U$.
It is composed of the intra- (inter-) orbital interaction $U$ ($U'$), 
and the exchange interaction $J=(U-U')/2$.
Below, we fix the ratio $J/U=0.1$.
The $4\times4$ matrix expression of on-site Coulomb interaction
at each site, ${\hat U}^{s(c)}$ for spin (charge) channel,
is explained in the SM A.
In the mean-field-level approximation,
the spin instability is the most prominent
because of the largest interaction $U$.
Figure \ref{fig:fig1} (e) exhibits the 
intra-$b_{3g}$-orbital static ($\w=0$) spin susceptibilities
$\chi^s_{A}(\q)\equiv \chi^s_{AA,AA}(\q)$ and 
$\chi^s_{\rm tot}(\q)=\sum_m^{A,B,C}\chi^s_{m}(\q)$ in the random phase approximation (RPA)
at $U=1.26$ ($\a_S=0.80$ at $T=0.02$).
In the RPA,
$\hat{\chi}^s(q)=\hat{\chi}^0(q)(\hat{1}-\hat{\Gamma}^s\hat{\chi}^0(q))^{-1}$,
where $\hat{\chi}^0(q)$ is the 
irreducible susceptibility matrix
and $q\equiv(\q,\w_l=2\pi Tl)$.
The spin Stoner factor $\a_S$
is the maximum eigenvalue of $\hat{\Gamma}^s\hat{\chi}^0(q)$,
and magnetic order appears when $\a_S=1$.
Thus, intra-orbital spin susceptibility gets enhanced
at $\q\approx \bm{0}$ in the present kagome model.
(Note that $\chi^s_{A}(\q_1)$ is small
because orbitals $A$ and $B$ correspond to different sites,
referred to as the sublattice interference \cite{Thomale2013}.
Also, $\chi^s_{AA,BB}, \chi^s_{A'}$ is much smaller than $\chi^s_{A}$.)

\subsection{bond-order derived from DW equation}

Nonmagnetic DW orders cannot be explained in the RPA 
unless large nearest-neighbor Coulomb interaction $V$ ($V>0.5U$) exists.
However, beyond-RPA nonlocal correlations,
called the vertex corrections (VCs), 
can induce various DW orders even for $V=0$
\cite{Kontani-PRB2011,Onari-SCVC,Onari-FeSe,Yamakawa-FeSe,Onari-AFN}.
To consider the VCs due to the paramagnon interference 
in Fig. \ref{fig:fig1} (a),
which causes the nematicity in Fe-based and cuprate superconductors,
we employ the linearized DW equation 
\cite{Onari-FeSe,Kawaguchi}: 
\begin{eqnarray}
\lambda_{\q}f_\q^{L}(k)&=& -\frac{T}{N}\sum_{p,M_1,M_2}
I_\q^{L,M_1}(k,p) 
\nonumber \\
& &\times \{ G(p)G(p+\q) \}^{M_1,M_2} f_\q^{M_2}(p) ,
\label{eqn:DWeq}
\end{eqnarray}
where $I_\q^{L,M}(k,p)$ is the 
``electron-hole pairing interaction'',
$k\equiv (\k,\e_n)$ and $p\equiv (\p,\e_m)$
($\e_n$, $\e_m$ are fermion Matsubara frequencies).
$L\equiv (l,l')$ and $M_i$ represent the pair of $d$-orbital indices
$A,B,C,A',B',C'$.
$\lambda_{\q}$ is the eigenvalue that represents the
instability of the DW at wavevector $\q$, and
$\max_\q\{\lambda_\q\}$ reaches unity at $T=T_{\rm DW}$.
$f_\q^L(k)$ is the Hermite form factor that is proportional to the 
particle-hole (p-h) condensation 
$\sum_\s \langle c_{\k+\q,l,\s}^\dagger c_{\k,l',\s}\rangle$,
or equivalently, the symmetry breaking component in the self-energy. 

It is important to use the appropriate kernel function $I_\q^{L,M}$,
which is given as 
$\delta^2 \Phi_{\rm LW}/\delta G_{l'l}(k)\delta G_{mm'}(p)$
at $\q={\bm 0}$ in the conserving approximation scheme 
\cite{BK,Onari-AFN},
where $\Phi_{\rm LW}$ is the Luttinger-Ward function introduced in the Method section.
If we apply the bare interaction to $I_\q^{L,M}$
that corresponds to RPA \cite{BK},
the relation $\lambda_\q>\a_S$ cannot be realized when $H_U$ is local.
Thus, higher-order corrections are indispensable.

Here, we apply the one-loop approximation for $\Phi_{\rm LW}$
\cite{Onari-SCVC,Onari-AFN}.
Then, $I_\q^{L,M}$ is
composed of one single-magnon exchange Maki-Thompson (MT) term
and two double-magnon interference 
Aslamazov-Larkin (AL) terms.
Their diagrammatic and analytic expressions 
are explained in the Method section.
Due to the AL terms,
the nonmagnetic nematic order in FeSe is naturally reproduced
even if spin fluctuations are very weak
\cite{Onari-SCVC}.
The importance of AL terms was
verified by the functional-renormalization-group (fRG) study
with constrained-RPA,
in which higher-order parquet VCs are produced 
in an unbiased way,
for several Hubbard models
\cite{Tazai-rev2021,Tsuchiizu1,Tsuchiizu4}.
Later, we see that the AL diagrams induce the 
backward and umklapp scattering shown in Fig. \ref{fig:fig1} (f),
and they mediate the p-h condensation 
at the inter-vHS nesting vector $\q_1=\k_{\rm B}-\k_{\rm A}$.

\begin{figure}[htb]
\includegraphics[width=.99\linewidth]{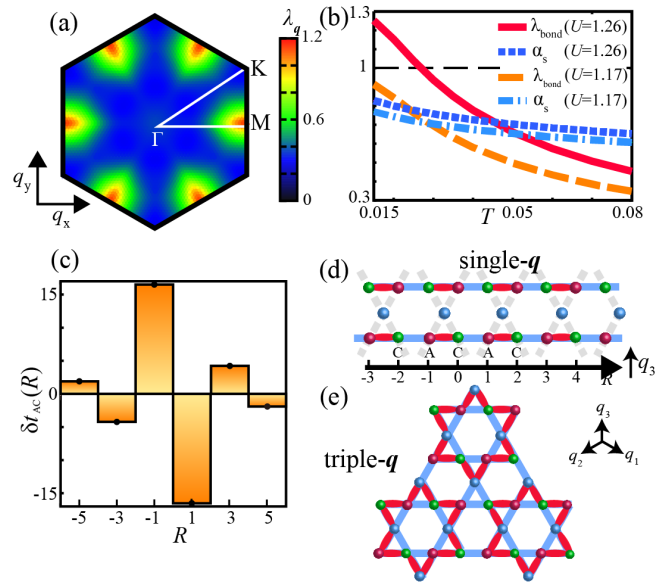}
\caption{
{\bf bond-order solution derived from DW equation:}
(a) Obtained $\q$-dependence of the eigenvalue $\lambda_\q$ at $n=3.8$
($T=0.02$ and $\a_S=0.80$).
$\lambda_\q$ shows peaks at $\q_n$ ($n=1,2,3$),
consistently with experiments in AV$_3$Sb$_5$.
(b) $T$-dependences of $\lambda_{\rm bond}$ and $\a_S$ 
at $U=1.26$ and $1.17$.
The DW susceptibility ($\chi^c_f(\q_n)\propto 1/(1-\lambda_{\rm bond})$)
increases as $T\rightarrow T_{\rm DW}\approx0.025$,
whereas magnetic susceptibility ($\chi^s({\bm0})\propto 1/(1-\a_S)$)
is almost constant.
(c) Modulation of hipping integrals 
$\delta t_{AC}(R{\bm a}_{\rm AC})$ for $\q=\q_3$ along the A-C direction
(arbitrary unit).
Its schematic picture at wavevector $\q_3$ is shown in (d).
(e) Expected triple-$\q$ star of David bond-order.
}
\label{fig:fig2}
\end{figure}

Figure \ref{fig:fig2}
(a) exhibits the obtained $\q$-dependence of the eigenvalue 
$\lambda_\q$ at $n=3.8$ ($T=0.02$ and $\a_S=0.80$).
The obtained peak position at $\q_n$ ($n=1,2,3$) 
is consistent with experiments in AV$_3$Sb$_5$.
The $T$-dependences of $\lambda_{\rm bond}\equiv \lambda_{\q_n}$ and $\a_S$ are 
shown in Fig. \ref{fig:fig2} (b).
The DW susceptibility ($\chi^c_f(\q_n)\propto 1/(1-\lambda_{\rm bond})$)
increases as $T\rightarrow T_{\rm DW}\approx0.025$,
whereas the increment of 
ferromagnetic susceptibility ($\chi^s({\bm0})\propto 1/(1-\a_S)$)
is small.
Then, what order parameter is obtained?
To answer this question, 
we perform the Fourier transform of the form factor:
\begin{eqnarray}
\delta t_{lm}(\r) &=& \frac1N \sum_{\k}f_{\q_n}^{lm}(\k) e^{i\r\cdot\k} .
\label{eqn:form-r}
\end{eqnarray}
Then, $\delta t_{lm}(\r_i-\r_j) \cos(\q_n\cdot \r_i +\theta)$
represents the modulation of the hopping integral
between $\r_i$ and $\r_j$,
where $\r_i$ represents the center of a unit-cell $i$
in real space, and $\theta$ is a phase factor.
The bond-order preserves the time-reversal-symmetry because it satisfies the relation 
$\delta t_{lm}(\r)=\delta t_{ml}(-\r)={\rm real}$.
(Note that the current order is
$\delta t_{lm}(\r)=-\delta t_{ml}(-\r)={\rm imaginary}$.)
Figure \ref{fig:fig2} (c) represents the obtained form factor
$\delta t_{AC}(\r)$ for $\q=\q_3$ along the A-C direction,
where ${\bm r}=R{\bm a}_{\rm AC}$ with odd number $R$.
The obtained solution is a bond-order because the relation
$\delta t_{CA}(\r)=\delta t_{AC}(-\r)$ is verified.
The relation $\delta t_{AA}(\r)=\delta t_{CC}(\r)=0$
holds in this bond-order solution.

To summarize,
we obtain the single-$\q$ smectic bond-order  
depicted in Fig. \ref{fig:fig2} (d).
In SM B,
we perform the DW equation analysis for $n=3.6$ and $3.7$
and obtain very similar results to Fig. \ref{fig:fig2}.
Thus, the robustness of the bond-order is confirmed,
irrespective of the Lifshitz transition at $n\approx3.71$.
In the triple-$\q$ state, in which 
three bond-orders with $\q_1$, $\q_2$, $\q_3$ coexist,
a star of David bond-order in Fig. \ref{fig:fig2} (e) appears.
Figure \ref{fig:figS4} in SM C
shows the unfolded FS
under the triple-$\q$ order below $T_{\rm DW}$.
In the present model, 
triple-$\q$ order is expected to emerge
because the momentum conservation 
$\q_1+\q_2+\q_3={\bm 0}$ gives rise to the third-order 
Ginzburg-Landau (GL) free energy
$F^{(3)}= b \phi_1\phi_2\phi_2$, where $\phi_n$ is real
order parameter for $\q=\q_n$ bond-order ($n=1-3$)  
\cite{Hirata}.
Here, $\phi_n {\hat f}_{\q_n}(k)$ is the bond-order function,
where ${\hat f}_{\q_n}(k)$ is the normalized dimensionless 
form factor given by the linearized DW equation.
A more detailed explanation is given in the SM D.

\begin{figure}[htb]
\includegraphics[width=.90\linewidth]{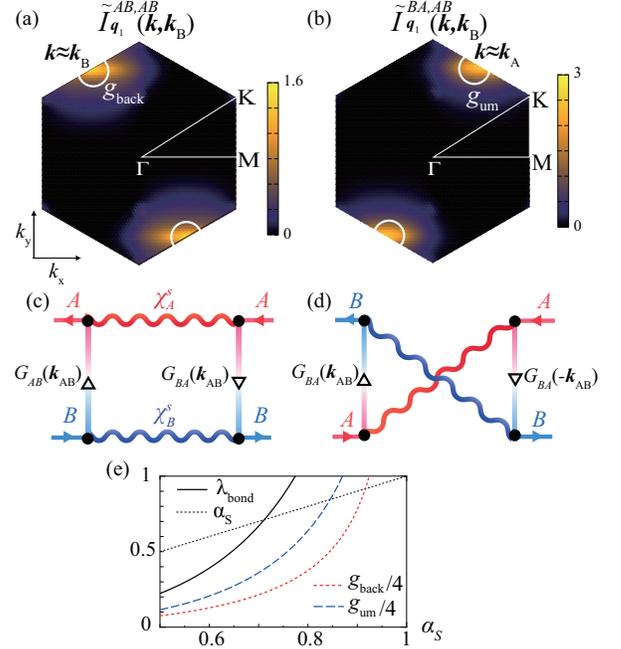}
\caption{
{\bf Origin of backward and umklapp scatterings that cause bond-order and SC state:}
Kernel function in the DW equation with orbital weights
at the lowest Matsubara frequency:
(a) ${\tilde I}_{\q_1}^{AB,AB}(\k,\k_{\rm B})$ and 
(b) ${\tilde I}_{\q_1}^{BA,AB}(\k,\k_{\rm B})$ for $\a_S=0.80$.
In the kernel function,
the outer momenta and sublattices are 
explained in the Method section.
The former at $\k=\k_{\rm B}$  and the latter at $\k=\k_{\rm A}$
give $g_{\rm back}$ and $g_{\rm um}$, respectively.
Both scatterings contribute to the bond-order formation.
(c) AL-VC with p-h pair and 
(d) that with p-p pair.
The former (latter) gives large $g_{\rm back}$ ($g_{\rm um}$).
(e) $\lambda_{\rm bond}$, $g_{\rm back}$ and $g_{\rm um}$
as functions of $\a_S$ at $T=0.02$.
}
\label{fig:fig3}
\end{figure}

We stress that the bond-order originates from the 
inter-sublattice VC in the kernel function $I$ 
in Eq. (\ref{eqn:DWeq}).
(Within the RPA, $I \ (=-U)$ is an intra-sublattice function.)
The dominant form factor at wavevector $\q=\q_1$, $f_{\q_1}^{lm}(\k)$,
is given by $(lm)=(AB)$ and $(BA)$.
To understand its origin, we examine the 
kernel function at the lowest Matsubara frequency,
multiplied with the $b_{3g}$-orbital weight ($A$, $B$, or $C$)
on two conduction bands at four outer points, ${\tilde I}$.
Results are shown in Figs. \ref{fig:fig3}
(a) ${\tilde I}_{\q_1}^{AB,AB}(\k,\k_B)$ and 
(b) ${\tilde I}_{\q_1}^{BA,AB}(\k,\k_B)$,
at $T=0.02$ and $\a_S=0.80$.
They are obtained in the triangular lattice model in Fig. \ref{fig:figS1} 
that is equivalent to the kagome metal.
We see the strong developments of
(a) $g_{\rm back}\equiv {\tilde I}_{\q_1}^{AB,AB}(\k_{\rm B},\k_{\rm B})$ and 
(b) $g_{\rm um}\equiv {\tilde I}_{\q_1}^{BA,AB}(\k_{\rm A},\k_{\rm B})$,
which correspond to the backward and umklapp scattering 
in Fig. \ref{fig:fig1} (f).
(We note the relation $\k_{\rm A}+\q_1=\k_{\rm B}$, and
four outer momenta and sublattices of $I$ are explained 
in the Method section.)
Both scatterings contribute to the bond-order formation, 
as we clearly explain based on a simple two vHS model
in SM B.
Microscopic origin of large $g_{\rm back}$ [$g_{\rm um}$] is the 
AL-VC with p-h [particle-particle (p-p)] pair 
shown in Fig. \ref{fig:fig3} (c) [(d)],
because of the relation $\chi^s_A,\chi^s_B\gg |\chi^s_{AA,BB}|$.
They are included as AL1 and AL2 in the kernel function $I$; see 
the Method section.

In Fig. \ref{fig:fig3} (e), we display the increment of 
$\lambda_{\rm bond}$, $g_{\rm back}$ and $g_{\rm um}$ 
with $\a_S \ (\propto U)$ at $T=0.02$.
(The relation $\lambda_{\rm bond}\propto g_{\rm back}+g_{\rm um}$ 
holds, as we explain in SM B.)
When $\a_S=0.75$, then $\lambda_{\rm bond}\approx 0.88$, 
$g_{\rm um}\approx 2$ and $g_{\rm back}\approx 1$, respectively.
Thus, both $g_{\rm um}$ and $g_{\rm back}$ are comparable or larger than $U$
due to the quantum interference mechanism in Figs. \ref{fig:fig3} (c) and (d),
in which the inter-orbital Green function $G_{AB}(k)$ is significant.
As understood from Fig. \ref{fig:fig1} (b),
$G_{AB}(k)$ is large at $\k\sim \k_{\rm AB}$,
while it vanishes at $\k=\k_{\rm A}$ and $\k_{\rm B}$.
Therefore, the FS portion away from vHS points
is indispensable in deriving the smectic order.


\begin{figure}[htb]
\includegraphics[width=.9\linewidth]{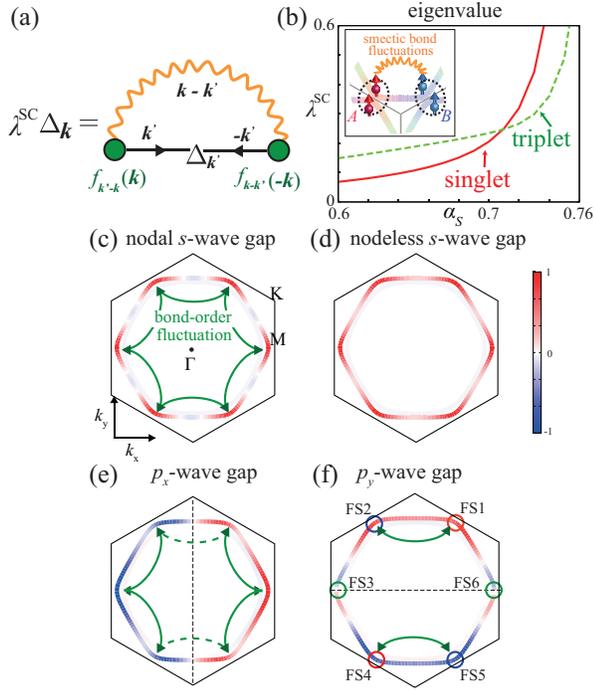}
\caption{
{\bf Unconventional SC states due to bond-order fluctuation ``beyond-Migdal'' pairing glue:}
(a) Pairing gap equation due to bond-order fluctuations.
The form factor $f$ gives the 
nonlocal (beyond-Migdal) electron-boson coupling function.
(b) Obtained eigenvalues of gap equation for 
the singlet $s$-wave ($A_{1g}$) and
the triplet $p$-wave ($E_{1u}$) states.
Obtained gap functions:
(c) nodal $s$-wave state ($\a_S=0.75$), 
(d) nodeless $s$-wave state ($\a_S=0.76$), and
(e)(f) $(p_x,p_y)$-wave state ($\a_S=0.70$).
Green full (broken) arrow lines represent the 
smectic fluctuations between vHS points 
with the same (opposite) sign gap functions.
}
\label{fig:fig4}
\end{figure}

\subsection{Unconventional superconductivity}

Finally, we study the unconventional superconductivity
mediated by bond-order fluctuations.
Here, we solve the following linearized SC gap equation on the FSs:
\begin{eqnarray}
\lambda^{\rm{SC}} \Delta_{\k}(\epsilon_{n})=
\frac{\pi T}{(2\pi)^2}\sum_{\epsilon_{m}}
\oint_{\rm FSs} \frac{d\k'}{v_{\k'}}
\frac{\Delta_{\k'}(\epsilon_{m})}{|\epsilon_{m}|} V^{\rm SC}_{\rm s(t)}(k,k') ,
\label{eqn:linear}
\end{eqnarray}
where $\Delta_{\k}(\epsilon_{n})$ is the gap function on FSs,
and $v_{\k}$ is the Fermi velocity. 
The eigenvalue $\lambda^{\rm{SC}}$ reaches unity at $T=T_{\rm c}$.
The diagrammatic expression of Eq. (\ref{eqn:linear}) is given in Fig. \ref{fig:fig4} (a).
The form factor represents the ``nonlocal'' electron-boson coupling function
that is a part of the beyond-Migdal effects.
$V^{\rm SC}_{\rm s/t}$ is the singlet/triplet pairing interaction in the band-basis,
due to the triple-$\q$ bond-order fluctuations ($V_{\rm bond}$)
and spin fluctuations ($\frac32 U^2\chi^s$)
derived in SM E.
Here, $V_{\rm bond}$ for $\k'-\k\approx \q_1$ is given as
\begin{eqnarray}
\frac12 \frac{g_{\rm um} {\bar f}_{\q_1}(\k) {\bar f}_{\q_1}(-\k')^*}
{1-\lambda_{\rm bond}}\frac{1}{1+\xi^2(\q_1-(\k'-\k))^2},
\label{eqn:pairV}
\end{eqnarray}
where 
${\bar f}_{\q}(\k)$ is the Hermite form factor in the band-basis,
and $|\bar{f}_{\q_1}(\k_{\rm A})|=1$.
Both $\lambda_{\rm bond}$ and $g_{\rm um}$ are already obtained in Fig. \ref{fig:fig3} (e),
and the numerator of Eq. (\ref{eqn:pairV}) on outer FS is
given in Fig. \ref{fig:figS6} in SM E.

Figure \ref{fig:fig4} (b) shows the obtained $\lambda^{\rm SC}$
at $T=0.02$ and $\xi=1.0$, where the $s$-wave singlet state appears 
when $\alpha_{S} \gtrsim 0.7$ and $\lambda_{\rm bond}>\alpha_{S}$.
Figures \ref{fig:fig4} (c) and (d) exhibit the 
obtained nodal $s$-wave gap function at $\a_S=0.75$ ($\lambda_{\rm bond}=0.88$)
and nodeless $s$-wave one at $\a_S=0.76$ ($\lambda_{\rm bond}=0.92$), respectively.
On the other hand, $(p_x, p_y)$-wave gap functions obtained 
at $\a_S=0.70$ are shown in Figs. \ref{fig:fig4} (e) and (f).
Note that obtained SC gap on inner FS made of $b_{2g}$-orbital is very small,
while it can be large due to (for instance) finite inter-band 
electron-phonon interaction.

Here, we discuss the origin of the $s/p$-wave SC state.
Triple-$\q$ bond-order fluctuations work as attraction between FS$i$ and FS($i+1$),
where FS$i$ ($i=1\sim6$) is the FS portion around vHS points shown in 
Fig. \ref{fig:fig4} (f).
Therefore, six pairs shown by green full arrows contribute to the 
$s$-wave state in Fig. \ref{fig:fig4} (c).
In contrast, only two pairs contribute to the $p_y$-wave state in Fig. \ref{fig:fig4} (f).
(In the $p_x$-wave state in Fig. \ref{fig:fig4} (e),
four pairs (two pairs) give a positive (negative) contribution.)
Therefore, the $s$-wave state is obtained for $\alpha_S\gtrsim0.7$,
where $\lambda_{\rm bond}$ exceeds $\a_S$.
In contrast, the $p$-wave state is obtained for $\alpha_S\lesssim0.7$,
because weak ferro-spin-fluctuations favor (destroy) the 
triplet (singlet) pairing.
Thus, the present spin + bond-order fluctuation mechanism 
leads to rich $s$- and $p$-wave states.
Possible SC states in the $P$-$T$ phase diagram in CsV$_3$Sb$_5$ 
will be discussed in the Discussion section.


The nodal gap structure shown in Fig. \ref{fig:fig4} (c) 
is obtained in the case of $\a_S = 0.75$ $(U = 1.18)$. 
We verified that the nodal $s$-wave gap structure emerges away from the 
vHS points so as to minimize the ``depairing effect by moderately 
$\k$-dependent repulsion by weak spin fluctuations'', 
which are shown in Fig. \ref{fig:fig1} (e). 
On the other hand, the nodeless $s$-wave state is realized when 
$\a_S = 0.76$, as shown in Fig. \ref{fig:fig4} (d).
The reason is that the attraction due to bond-order susceptibility 
($\propto 1/(1-\lambda_{\rm bond})$) is strongly enhanced with 
increasing for $\a_S \gtrsim 0.7$ as recognized in 
Fig. \ref{fig:fig3} (e), and therefore the reduction of depairing 
due to nodal structure becomes unnecessary.

To summarize, large attraction between different vHS points 
is induced by the bond-order fluctuations due to 
the paramagnon interference process.
In contrast, the repulsion between different vHS points 
due to spin fluctuations is small, 
by reflecting the fact that the vHS points $\k_A$, $\k_B$, and $\k_C$ 
are respectively composed of single orbital A, B, and C
(= sublattice interference \cite{Thomale2013}).
For this reason, moderate bond-order fluctuations 
($\lambda_{\rm bond}\gtrsim0.9$) can induce nodeless $s$-wave SC gap state 
against spin fluctuations.






\section{Discussion}
\label{sec:Discussion}

\subsection{Importance of paramagnon interference}
We have studied the exotic density-wave 
and beyond-Migdal unconventional superconductivity 
in kagome metal AV$_3$Sb$_5$ (A=K, Rb, Cs),
by focusing on the paramagnon interference mechanism.
This beyond-mean-field mechanism provides sizable 
``inter-sublattice'' scattering,
and therefore the smectic bond-order is realized 
in the presence of experimentally observed weak spin fluctuations.
The bond-order fluctuations naturally mediate strong 
pairing glue that leads to the $s$-wave state,
consistently with recent several experiments
\cite{kagome-full,kagome-full2,NMR2,impurity}.
Thus, the origins of the star of David order, the exotic superconductivity,
and the strong interplay among them,
are uniquely explained based on the paramagnon interference mechanism.
This mechanism has been overlooked previously.
This novel mechanism overcomes the difficulty 
of the sublattice interference \cite{Thomale2013}, 
which leads to tiny inter-site interaction in weak-coupling theories,
and gives rise to rich phase transitions in kagome metals.
These key findings would promote future experiments on not only kagome metals,
but also other frustrated metals.

A great merit of the present theory is that
the bond-order is robustly obtained for a wide range of model parameters,
as long as the bandstructure near the three vHS points is 
correctly reproduced.
$U$ is the only model parameter in the present theory.
To clarify this merit,
we make the comparison between the DW equation theory and mean-field theory.
In the mean-field theory, the instability of the 
charge bond-order is always secondary even if large nearest-neighbor Coulomb interaction
$V$ is introduced.
In contrast, in the DW equation theory,
the charge bond-order solution is robustly obtained even when $V=0$.
This is a great merit of the present DW equation analysis.
This merit remains even
if both charge- and spin-channel VCs are taken into account as explained 
in SM F.

We also discuss interesting similarities 
between kagome metal and other strongly correlated metals.
The paramagnon interference mechanism
has been successfully applied to explain
the nematic and smectic orders
in Fe-based and cuprate superconductors
\cite{Onari-SCVC,Tazai-rev2021}.
However, they appear only in the vicinity of the magnetic criticality,
except for FeSe systems
\cite{Onari-FeSe,Yamakawa-FeSe}.
In contrast, the smectic bond-order in kagome metal
appears irrespective of small spin fluctuations ($\a_S\sim0.75$),
because of the strong geometrical frustration inherent in kagome metals.
The present study would be useful to understand the 
recently discovered ``smectic order and adjacent high-$T_{\rm c}$ state''
in FeSe/SrTiO$_3$ \cite{smectic-FeSe}.


\subsection{Impurity effect on superconductivity}

The impurity effect is one of the most significant experiments
to distinguish the symmetry of the SC gap function.
However, experimental reports of the impurity effect on AV$_3$Sb$_5$ 
and its theoretical analysis have been limited so far.
Here, we study the nonmagnetic impurity effect on both $s$-wave and $p$-wave SC states
predicted in the present theory in Fig. \ref{fig:fig4}.
We treat the dilute V-site impurities based on the T-matrix approximation.
The impurity potential on the A-site is $(\hat{I}_{\rm imp}^{\rm A})_{ll'}=\delta_{l,l'}$,
where $l,l'=A,A'$.
In this case, the T-matrix on A-site is given by 
$\hat{T}^{\rm A}=\hat{I}_{\rm imp}^{\rm A} (\hat{1}-\hat{g}^{\rm A}\hat{I}^{\rm A}_{\rm imp})^{-1}$, 
where $\hat{g}^{\rm A}$ is the $2\times 2$ local Green function on A-site.
In this case, the normal self-energy is given by
$\hat{\Sigma}^{n}=n_{\rm imp}(\hat{T}^{\rm A}+\hat{T}^{\rm B}+\hat{T}^{\rm C})$,
where $n_{\rm imp}$ is the impurity concentration.
The anomalous self-energy is also given by the T-matrix.
Here, we consider the unitary limit case ($I_{\rm imp}=\infty$).
More detailed explanation is written in Ref. \cite{impurity-Onari}.

\begin{figure}[htb]
\includegraphics[width=.87\linewidth]{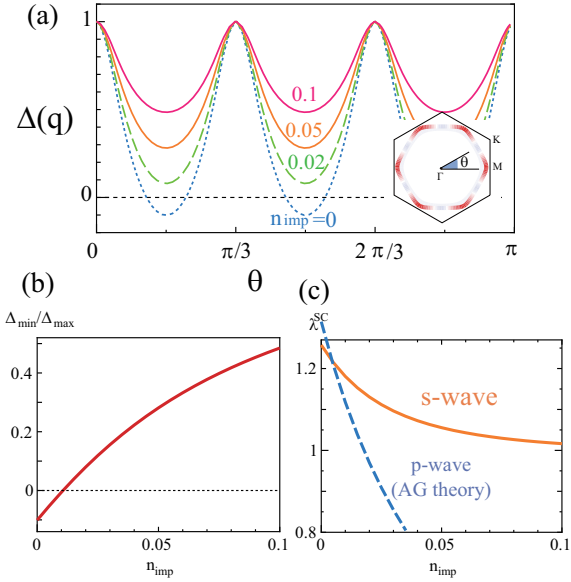}
\caption{
{\bf Impurity effect on superconductivity:}
(a) Obtained nodal $s$-wave gap function at $n_{\rm imp}=0-0.1$.
(b) $n_{\rm imp}$-dependence of $\Delta_{\rm min}/\Delta_{\rm max}$ 
in the $s$-wave state.
(c) $n_{\rm imp}$-dependence of the eigenvalue of $s$-wave and $p$-wave SC states.
$s$-wave superconductivity is robust against the impurity effect, 
while the $p$-wave one is quite weak.
(Here, $p$-wave pairing interaction is magnified by $2.7$.)
}
\label{fig:fignimp}
\end{figure}

Figure \ref{fig:fignimp} (a) shows the changes of the nodal $s$-wave SC gap function 
due to the impurity effect at $\alpha_{S}=0.75$.
The gap function at $n_{\rm imp}=0$ is the same as Fig. \ref{fig:fig4} (c).
The accidental nodes at $n_{\rm imp}=0$ are lifted up due to the impurities, 
and the nodeless $s$-wave gap emerges at just $n_{\rm imp}=0.02$. 
The ratio of the minimum gap over the maximum one quickly increases 
with $n_{\rm imp}$ as plotted in Fig. \ref{fig:fignimp} (b).
Figure \ref{fig:fignimp} (c) shows the eigenvalues of the
$s$-wave ($\lambda^{\rm SC}_{s}$) and $p$-wave SC states ($\lambda^{\rm SC}_{p}$).
Note that $\lambda^{\rm SC}_{s(p)}$ is proportional to $s(p)$-wave $T_{\rm c}$. 
(Here, the pairing interaction for the $p$-wave SC is magnified by $2.7$ 
to make both $\lambda^{\rm SC}_{s}$ and $\lambda^{\rm SC}_{p}$ comparable.)
$\lambda^{\rm SC}_{p}$ drastically decreases
with $n_{\rm imp}$ by following the Abrikosov-Gorkov theory.
In contrast, the reduction in $\lambda^{\rm SC}_{s}$ is much slower,
and its suppression saturates when the gap becomes nearly isotropic
for $n_{\rm imp} \gtrsim 0.05$.

The obtained impurity-induced drastic change in the gap anisotropy 
is a hallmark of the $s$-wave SC 
mediated by the bond-order fluctuations.
Thus, measurements of the impurity effects will be very promising 
toward the whole understanding of the SC phase. 
Note that when the $p$-wave SC state appears at $n_{\rm imp}=0$, 
the transition from $p$-wave to $s$-wave state is
caused by introducing dilute impurities.

\subsection{$P$-$T$ phase diagram}
\label{sec:P-T}

We discuss the $P$-$T$ phase diagram of CsV$_3$Sb$_5$,
in which the SC phase shows the highest $T_{\rm c} \sim 8$K 
at the critical pressure $P_{c2}\sim 2$GPa ($T_{\rm DW}=0$) \cite{kagome-P-Tc1}.
Inside the bond-order phase,
the second highest SC dome with $T_{\rm c}\sim6$K emerges at $P_{c1}\sim0.7$GPa. 
Between $P_{c1}$ and $P_{c2}$, both $T_{\rm c}$ and the SC volume fraction 
are reduced, while the residual resistivity increases.
As discussed in Ref. \cite{kagome-P-Tc1},
these states remind us of the inhomogeneous 
``nearly-commensurate CDW (NCCDW)'' in 1T-TaS$_2$,
which is realized when the correlation-driven 
incommensurate DW order at the FS nesting vector \cite{Hirata}
is partially locked to the lattice via the electron-phonon interaction. 
When such an inhomogeneous DW state appears,
$T_{\rm c}$ of strongly anisotropic SC gap state
should be suppressed, so the double-dome SC structure is realized.

\begin{figure}[htb]
\includegraphics[width=.80\linewidth]{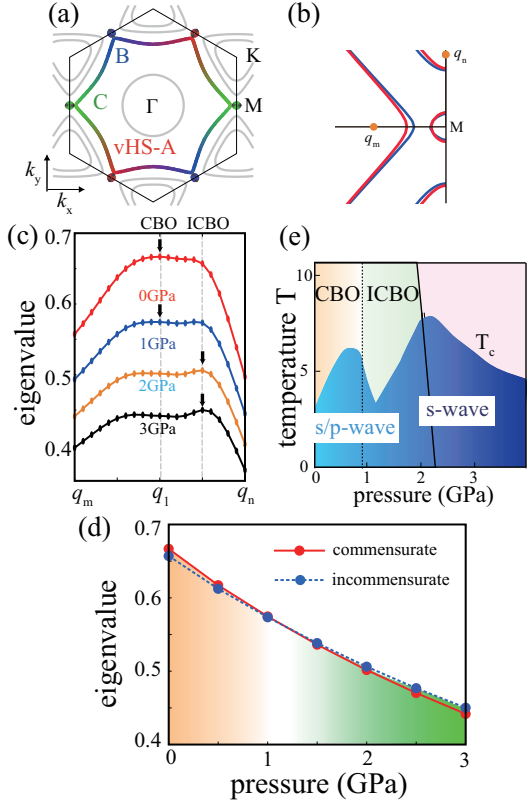}
\caption{
{\bf Pressure-induced C-IC bond-order transition:}
(a) FSs in the realistic 30 orbital model at $P=0$.
The $b_{3g}$-orbital weight on A (red), B (blue), and C (green) 
sublattices are shown.
(b) FSs around vHS points at $P=0$ and $3$GPa.
(c) Obtained $\q$-dependence of the eigenvalue for bond-order at $0\sim3$GPa. 
Here, CBO (ICBO) means the commensurate (incommensurate) bond-order.
(d) Pressure dependence of the eigenvalue of the bond-order. 
The C-IC transition emerges around $P\sim 1$GPa.
(e) Schematic $P$-$T$ phase diagram derived from the present theory.
}
\label{fig:figPT}
\end{figure}

To support this NCCDW scenario for $P>P_{c1}$ 
\cite{kagome-P-Tc1},
we construct realistic tight-binding models at $0-3$GPa
based on the first-principles study,
which are constructed by using Wien2k and Wannier90 software \cite{1stprin}. 
Figure \ref{fig:figPT} (a) shows the FSs at $P=0$: 
The $b_{3g}$-orbital FS is essentially similar to that in 
Fig. \ref{fig:fig1} (d).
Interestingly, the $b_{3g}$-FS at $3$GPa becomes smaller 
due to the pressure-induced self-doping on $b_{3g}$-FS ($\sim1.5$\%),
deviating from the vHS points as illustrated in Fig. \ref{fig:figPT} (b).
 (The change in $k_{\rm F}$ on $k_x$-axis is $\Delta k_{\rm F}=-0.02\pi$.)
The obtained change is reliable because 
it is derived from the first-principles 
``pressure Hamiltonian $\Delta H_0^{\rm DFT}(P)$'' given in the SM G.
The present discovered $P$-dependence in the FS and its nesting vector
would cause the C-IC bond-order transition.

Based on the derived realistic models,
we perform the DW equation analysis.
Figure \ref{fig:figPT} (c) shows the obtained $\q$-dependent 
eigenvalue, $\lambda_\q$, 
at $0-3$GPa with $T=0.04$ [eV] and $U=2.7$ [eV].
At $P=0$,
we obtain the commensurate bond-order (CBO) solution at $\q=\q_1$,
so the robustness of the bond-order solution 
in Fig. \ref{fig:fig2} is confirmed.
On the other hand,
$\lambda_{\q_1}$ is quickly suppressed under pressure (over 30\% at 3GPa).
Since $T_{\rm DW}\propto \lambda_{\q_1}$ qualitatively,
this result is consistent with the strong suppression of 
the bond-order under pressure in kagome metals.
In the interference mechanism, small reduction in $\alpha_{S}$
induced by the pressure (just $\sim0.03$ at 3GPa)
causes sizable suppression of $\lambda_{\q_1}$, 
as we can seen in Fig. \ref{fig:fig2} (b).
Interestingly, the CBO at $P=0$
turns to be incommensurate one at $\q=\q_1+(0,\delta)$ when $P\gtrsim1$GPa,
by reflecting the change in the nesting condition.
The $P$-dependence of the bond-order eigenvalues is
summarized in Fig. \ref{fig:figPT} (d).

In SM B,
we examine the filling-dependence of the bond-order solution
in the present six orbital kagome lattice model.
As shown in Fig. \ref{fig:figS3-2} (b),
the C-IC bond-order transition occurs at $n=n_0\equiv 3.82$.
For $n>n_0$, the incommensurate bond-order (ICBO) is realized 
due to the change in the Fermi momentum $\Delta k_{\rm F}$.
Thus, the C-IC transition can also be understood
in the present simple six orbital model.
The present theory supports the NCCDW scenario 
discussed in Ref. \cite{kagome-P-Tc1}.

Next, we propose a possible scenario 
for the double-dome SC phase on AV$_3$Sb$_5$.
The phase diagram based on the present scenario
is schematically represented in Fig.\ref{fig:figPT} (e).
The present bond-order fluctuation-mediated $s$-wave state
should exhibit the highest-$T_{\rm c}$ around the critical pressure $P=P_{c2}$.
Thus, the $T_{\rm c}$ monotonically decreases as $|P-P_{c2}|$ increases. 
In addition, the NCCDW-like inhomogeneous states
triggered by the ICBO formation lead to the dip structure
in $T_{\rm c}$ for $P\gtrsim P_{c1}$.
Therefore, the double-dome SC phase is naturally explained in 
terms of the C-IC bond-order transition.

In another SC dome for $P<P_{c1}$, 
both $p$- and $s$-wave SC can emerge
because bond-order and spin fluctuations would be comparable. 
If the $p$-wave SC state is realized at $n_{\rm imp}=0$, 
the $p$-wave to $s$-wave SC transition will occur
at $n_{\rm imp}\sim 0.01$ as understood in Fig. \ref{fig:fignimp} (c).

\subsection{Future problems}


In kagome metals,
the bond-order state is the platform of various exotic phenomena.
In this respect, the mechanism of the bond-order state 
should be clarified in kagome metal.
The discovered quantum interference process in the present study 
triggers the bond-order formation, 
and this process would be important even below $T_{\rm DW}$.
Thus, the present study paved the way for understanding the whole phase diagram.


A central open problem in the bond-order state
is the time-reversal-symmetry-breaking (TRSB) state.
In AV$_3$Sb$_5$, the TRSB state has been reported by STM, 
Kerr effect and $\mu$SR measurements 
in Refs. \cite{STM1,mSR_TRS,Kerr}.
The $T_{\rm TRSB} \sim 70$K is suggested by $\mu$SR study, while 
$T_{\rm TRSB}=T_{\rm DW}=94$K is reported by Kerr effect  \cite{STM1,Kerr}.
The leading candidate for the TRSB is the charge LC order that 
accompanies the local magnetic field,
as studied in cuprates for a long time \cite{Varma,Affleck}.

However, the microscopic mechanism of the LC order has been unsolved.
For example, the LC phase does not appear in the $U$-$V$ phase diagram
in the mean-field approximation in Fig. \ref{fig:figS7}.
Thus, beyond-mean-field analysis is required to solve this open issue.
An important clue is given by the 
spin-fluctuation-driven LC mechanism in frustrated metals
in Refs. \cite{Kontani-sLC,Tazai-cLC}.
This beyond-mean-field LC mechanism is general 
because the LC is caused by various spin/charge-channel fluctuations.
Thus, new spin/charge-channel fluctuations
due to the FS reconstruction below $T_{\rm DW}$ may induce the LC order 
in the bond-order state.
Therefore, the present bond-order theory provides a significant starting 
point to understand the phase diagram of AV$_3$Sb$_5$.

The coexistence of the LC and the bond-order 
is predicted based on GL theory in Ref. \cite{Lin2021}. 
Interestingly, the relation $T_{\rm TRSB} \sim T_{\rm DW}$ is realized when 
the third-order term in the GL free energy, inherent in kagome metals, is sizable.
In future, it is useful to solve the ``full DW equation without linearization'', 
in which effect of the third-order GL term is included.

Another important issue is the anomalous transport phenomena below $T_{\rm DW}$.
For instance, giant anomalous Hall effect \cite{AHE1,AHE2} is observed 
in several kagome metals.
In addition, sizable thermoelectric power and Nernst effect 
are reported \cite{Nernst}.
These transport coefficients can be calculated based on the 
realistic tight-binding models in Fig. \ref{fig:figPT},
under the presence of the bond-order and the LC order.
The VCs for the current due to spin/charge fluctuations would
play significant roles \cite{Kontani-ROP}.
It is an useful future problem to study the effect of the 
three-dimensionality on the electronic states in kagome metals.



\section{Materials and Methods}
\label{sec:Methods}
\subsection{Derivation of density-wave equation}

Here, we derive the kernel function in the
DW equation, $I^{l l', m m'}_{\q}(k,k')$,
studied in the main text.
It is given as 
$\delta^2 \Phi_{\rm LW}/\delta G_{l'l}(k)\delta G_{mm'}(p)$
at $\q={\bm 0}$ in the conserving approximation scheme 
\cite{BK,Onari-AFN},
where $\Phi_{\rm LW}$ is the Luttinger-Ward function.
Here, we apply the one-loop approximation for $\Phi_{\rm LW}$
\cite{Onari-SCVC,Onari-AFN}.
Then, $I_\q^{L,M}$ in this kagome model is given as
\begin{eqnarray}
&& I^{l l', m m'}_{\q} (k,k')
= \sum_{b = s, c} \frac{a^b}{2} 
\Bigl[ -V^{b}_{l m, l' m'} (k-k') 
\nonumber \\
&&
+\frac{T}{N} \sum_{p} \sum_{l_1 l_2, m_1 m_2}
	V^{b}_{l l_1, m m_1} \left( p+\q \right) 
	V^{b}_{m' m_2, l' l_2} \left( p \right)
 \nonumber \\
&& \qquad\qquad\qquad \quad
\times G_{l_1 l_2} (k-p) G_{m_2 m_1} (k'-p)
 \nonumber \\
&& 
+\frac{T}{N} \sum_{p} \sum_{l_1 l_2, m_1 m_2}
	V^{b}_{l l_1, m_2 m'} \left( p+\q \right) 
	V^{b}_{m_1 m, l' l_2} \left( p \right)
 \nonumber \\
&& \qquad\qquad\qquad 
\times G_{l_1 l_2} (k-p) G_{m_2 m_1} (k'+p+\q) \Bigr] ,
\label{eqs:kernel} 
\end{eqnarray}
where $a^{s(c)} = 3$($1$) and $p = (\p,\w_l)$. 
$\hat{V}^{b}$ is the $b$-channel interaction given by 
$\hat{V}^{b} = \hat{U}^{b} + \hat{U}^{b} \hat{\chi}^{b} \hat{U}^{b}$. 
$\hat{U}^{b}$ is the matrix expression of the 
bare multiorbital Coulomb interaction for channel $b$.

Under the uniform ($\q={\bm0}$) DW state,
the one-loop $\Phi_{\rm LW}$ is given as
$\displaystyle \Phi_{\rm LW}=T\sum_p [
\frac32{\rm Tr}\ln ({\hat 1}-{\hat U}^s{\hat \chi}^0(p))
+\frac12{\rm Tr}\ln ({\hat 1}-{\hat U}^c{\hat \chi}^0(p))]$
with the correction terms up to $O(U^2)$.
When the wavevector $\q$ of the DW state is nonzero,
${\hat \chi}^0(p)$ is replaced with ${\hat \chi}^0(p;q)$.

\begin{figure}[htb]
\includegraphics[width=.99\linewidth]{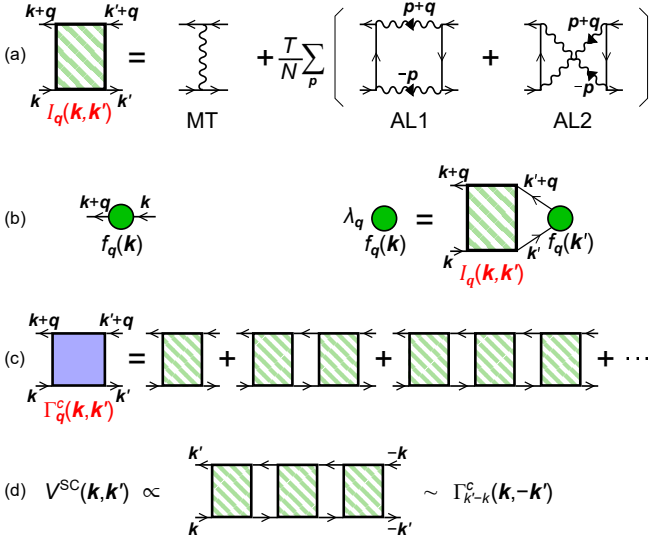}
\caption{
{\bf Derivations of DW equation and beyond-Migdal pairing interaction:}
(a) Charge-channel kernel function $I^{l l', m m'}_{\q} (k,k')$
(b) Linearized DW equation.
(c) Charge-channel full four-point vertex $\Gamma_\q^c(k,k')$ 
obtained by solving the DW equation.
(d) Pairing interaction due to $\Gamma_\q^c$:
$V^{\rm SC}(k,k') \propto \Gamma_{k'-k}^c(k,-k')$. 
}
\label{fig:figS2}
\end{figure}

The first term of Eq. (\ref{eqs:kernel}) 
corresponds to the single-magnon exchange Maki-Thompson term, 
and the second and third terms give two double-magnon interference AL terms. 
They are expressed in Fig. \ref{fig:figS2} (a).

The DW instability driven by 
nonlocal beyond-mean-field correlation ${\hat I}_{\q} (k,k')$
is obtained by solving the DW equation introduced in 
Refs. \cite{Onari-FeSe,Kawaguchi,Onari-AFN}:
\begin{eqnarray}
\lambda_\q f_\q^{ll'}(k)&=& \frac{T}{N}
\sum_{k',m,m'} {K}_{\bm{q}}^{ll',mm'}(k,k') f_\q^{mm'}(k'),
\label{eqn:linearized-DW} \\
{K}_{\bm{q}}^{ll',mm'}(k,k')&=& 
-\sum_{m'',m'''}{I}_{\bm{q}}^{ll',m'',m'''}(k,k')
\nonumber \\
& &\times G_{m''m}(k'+\q)G_{m'm'''}(k') ,
\label{eqn:linearized-DW2}    
\end{eqnarray}
which is depicted in Fig. \ref{fig:figS2} (b).
Here, $\lambda_\q$ is the eigenvalue
that reaches unity at the transition temperature.
$\hat{f}_\q$ is the form factor of the DW order,
which corresponds to the ``symmetry-breaking in the self-energy''.
By solving Eq. (\ref{eqn:linearized-DW}), we can obtain the optimized 
momentum and orbital dependences of $\hat{f}$.
This mechanism has been successfully applied to 
explain the electronic nematic orders in Fe-based
\cite{Onari-SCVC,Yamakawa-FeSe,Onari-FeSe} 
and cuprate superconductors \cite{Tazai-rev2021},
and multipole orders in $f$-electron systems
\cite{Tazai-CeB6}.

An arbitrary phase factor $e^{i\a}$ can be multiplied
to the solution of the linearized DW equation ${\hat f}_\q(k)$.
However, the phase factor should be determined uniquely 
so that ${\tilde f}_\q(\k)=({\hat f}_\q(\k,\pi T)+{\hat f}_\q(\k,-\pi T))/2$
satisfies the Hermite condition 
${\tilde f}^{lm}_\q(\k)= [{\tilde f}^{ml}_{-\q}(\k+\q)]^*$.

Finally, we discuss the effective interaction driven by 
the bond-order fluctuations.
By solving the DW equation (\ref{eqn:linearized-DW}), we obtain the 
full four-point vertex function $\Gamma^{c}_\q(k,k')$
that is composed of $I_\q^{c}$ and $G(k+\q)G(k)$
shown in Fig. \ref{fig:figS2} (c),
which increases in proportion to $(1-\lambda_\q)^{-1}$.
Thus, we obtain the relation
$\Gamma^{c}_\q(k,k')\approx f_\q(k)\{f_\q(k')\}^* {\bar I}_\q^c(1-\lambda_\q)^{-1}$,
which is well satisfied when $\lambda_\q$ is close to unity.

As we will discuss in SM E,
the pairing interaction due to the bond-order fluctuations is
given by the full four-point vertex:
$V^{\rm SC}(\k,\k')\sim \Gamma_{\k'-\k}^c(\k,-\k')
\sim f_{\k'-\k}(\k)\{f_{\k'-\k}(-\k')\}^*(1-\lambda_\q)^{-1}$,
which is depicted in Fig. \ref{fig:figS2} (d).

It is noteworthy that 
both the DW equation and the fRG method
explain the nematic and smectic bond-order in single-orbital 
square lattice Hubbard models \cite{Tsuchiizu4,Kawaguchi}
and anisotropic triangular lattice ones
\cite{Tazai-rev2021}.
This fact means that higher-order diagrams other than MT or AL terms,
that are included in the fRG method, are not essential
in explaining the bond-order.
Note that the contributions away from the conduction bands 
are included into $N$-patch fRG 
by applying the RG+cRPA method 
\cite{Tsuchiizu4,Tazai-rev2021,Tazai-rev2021}.


\subsection{Acknowledgments}

{\bf Funding}: 
This study has been supported by Grants-in-Aid for Scientific
Research from MEXT of Japan (JP18H01175, JP17K05543, JP20K03858, JP20K22328),
and by the Quantum Liquid Crystal
No. JP19H05825 KAKENHI on Innovative Areas from JSPS of Japan.

{\bf Author contributions}: R.T. performed the all calculations
discussing with Y.Y., S.O., and H.K., and R.T. and H.K. wrote the paper. 

{\bf Competing interests}: The authors declare that they have no competing interests.

{\bf Data and materials availability}: All data needed to evaluate the conclusions in the paper are present in the paper and/or the Supplementary Materials.


\clearpage
\newpage


\makeatletter
\renewcommand{\thefigure}{S\arabic{figure}}
\renewcommand{\theequation}{S\arabic{equation}}
\makeatother
\setcounter{figure}{0}
\setcounter{equation}{0}
\setcounter{page}{1}
\setcounter{section}{1}

\begin{widetext}
\begin{center}
{\bf \large 
[Supplementary Materials] \\
\vspace{3mm}
{\large
Mechanism of exotic density-wave 
and beyond-Migdal unconventional superconductivity 
in kagome metal AV$_3$Sb$_5$ (A=K, Rb, Cs)
}
}
\end{center}

\begin{center}
Rina Tazai, Youichi Yamakawa, Seiichiro Onari, and Hiroshi Kontani
\end{center}

\begin{center}
\textit{Department of Physics, Nagoya University, Nagoya 464-8602, Japan}
\end{center}

\end{widetext}

\subsection{A: Model Hamiltonian and RPA}

\begin{figure}[htb]
\includegraphics[width=.7\linewidth]{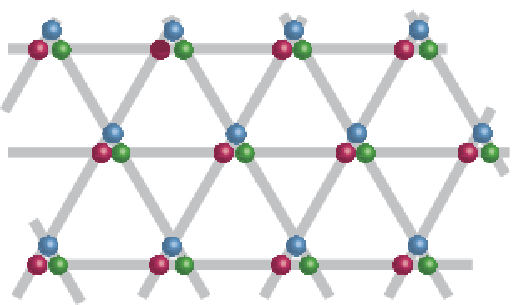}
\caption{
{\bf Six-orbital triangular lattice model:}
In this model, three apical sites (A, B, C) 
of each upper triangular site in the kagome lattice 
are located at the same position.
This model is convenient for the numerical study
because both intra- and inter-orbital susceptibilities
become periodic in the first BZ.
}
\label{fig:figS1}
\end{figure}

In the main text, we analyze the kagome lattice model
shown in Fig. \ref{fig:fig1} (b) 
introduced in Ref. \cite{Thomale2021}.
In this model, a unit cell contains three sites (A, B, C),
and each site possesses two orbitals ($b_{3g}$ and $b_{2g}$).
In the theoretical analysis, it is more convenient to
study a completely equivalent 
``six-orbital triangular lattice model'' in Fig. \ref{fig:figS1}:
It is derived from the kagome lattice model 
by shifting three apical sites (A, B, C)
of each upper triangular to its center,
without changing the hopping integrals and the Coulomb interaction terms.
One of the great merits of analyzing this triangular model is that
any inter-site vector ${\bm r}_i-{\bm r}_j$ is equal to 
a translation vector, and therefore functions in the momentum space 
(such as $\chi^s_{ll',mm'}(\q)$ and $f^{lm}_\q(\k)$) 
become periodic in the first Brillouin zone (BZ).
For this reason,
we perform the numerical study in the main text
based on the triangular lattice model in Fig. \ref{fig:figS1}.

Next, we explain the multiorbital Coulomb interaction.
The matrix expression of the 
spin-channel Coulomb interaction is
\begin{equation}
U_{l_{1}l_{2},l_{3}l_{4}}^s = \begin{cases}
U, & l_1=l_2=l_3=l_4 \\
U' , & l_1=l_3 \neq l_2=l_4 \\
J, & l_1=l_2 \neq l_3=l_4 \\
J' , & l_1=l_4 \neq l_2=l_3
\end{cases}
\end{equation}
in the case that 
$l_1 \sim l_4$ are orbitals ($X$, $X'$) at site X (=A,B,C).
In other cases, $U_{l_{1}l_{2},l_{3}l_{4}}^s =0$.
Also, the matrix expression of the 
charge-channel Coulomb interaction is
\begin{equation}
U_{l_{1}l_{2},l_{3}l_{4}}^c = \begin{cases}
-U, & l_1=l_2=l_3=l_4 \\
U'-2J , & l_1=l_3 \neq l_2=l_4 \\
-2U' + J, & l_1=l_2 \neq l_3=l_4 \\
-J' , & l_1=l_4 \neq l_2=l_3
\end{cases}
\end{equation}
in the case that 
$l_1 \sim l_4$ are orbitals ($X$, $X'$) at site X (=A,B,C).
In other cases, $U_{l_{1}l_{2},l_{3}l_{4}}^c =0$.
Here, $U$ ($U'$) is the intra-orbital (inter-orbital)
Coulomb interaction, $J$ is the Hund's coupling, 
and $J'$ is the pair hopping term.
In the main text, we assume the relations $U=U'+2J$ and $J=J'$,
and set the constraint $J/U=0.10$.
The obtained results are not sensitive to the ratio $J/U$.

The spin (charge) susceptibility in the RPA,
$\chi^{s(c)}_{ll',mm'}(q)$, is given by
\begin{eqnarray}
\hat{\chi}^{s(c)}(q)= \hat{\chi}^0(q)
(\hat{1}-\hat{U}^{s(c)}\hat{\chi}^0(q))^{-1} ,
\end{eqnarray}
where the element of the irreducible susceptibility is
$\chi^0_{ll',mm'}(q)=-\frac{T}{N}\sum_k G_{lm}(k+q)G_{m'l'}(k)$.
$G_{lm}(k)$ is the $(l,m)$ element of the electron Green function:
${\hat G}=(\e_n{\hat 1}-{\hat H}_0(\k))^{-1}$.

In the present model, 
$\chi^s_{ll',mm'}(q)$ is small unless all orbitals belong to $b_{3g}$.
Also, $\chi^s_{ll',mm'}(q)$ becomes large only when $l=l'=m=m'$
and $l=A$ or $B$ or $C$.
The spin susceptibility in the present model is shown in 
Fig. \ref{fig:fig1} (e) in the main text.

\subsection{B: Robustness of bond-order solution in the DW equation}

In the main text, we present the numerical results for $n=3.8$.
In this case, the $b_{3g}$-orbital FS is very close to 
the vHS points on the BZ boundary as shown in Fig.\ref{fig:fig1} (d),
consistently with recent ARPES reports.
In this model, the van Hove filling is $n_{\rm vHS}=3.71$, and 
the single large FS around $\Gamma$ point is divided into two pockets
around $K$ and $K'$ points for $n<n_{\rm vHS}$.
Thus, it is important to verify the robustness of numerical results 
for different electron filling $n$.

\begin{figure}[htb]
\includegraphics[width=.8\linewidth]{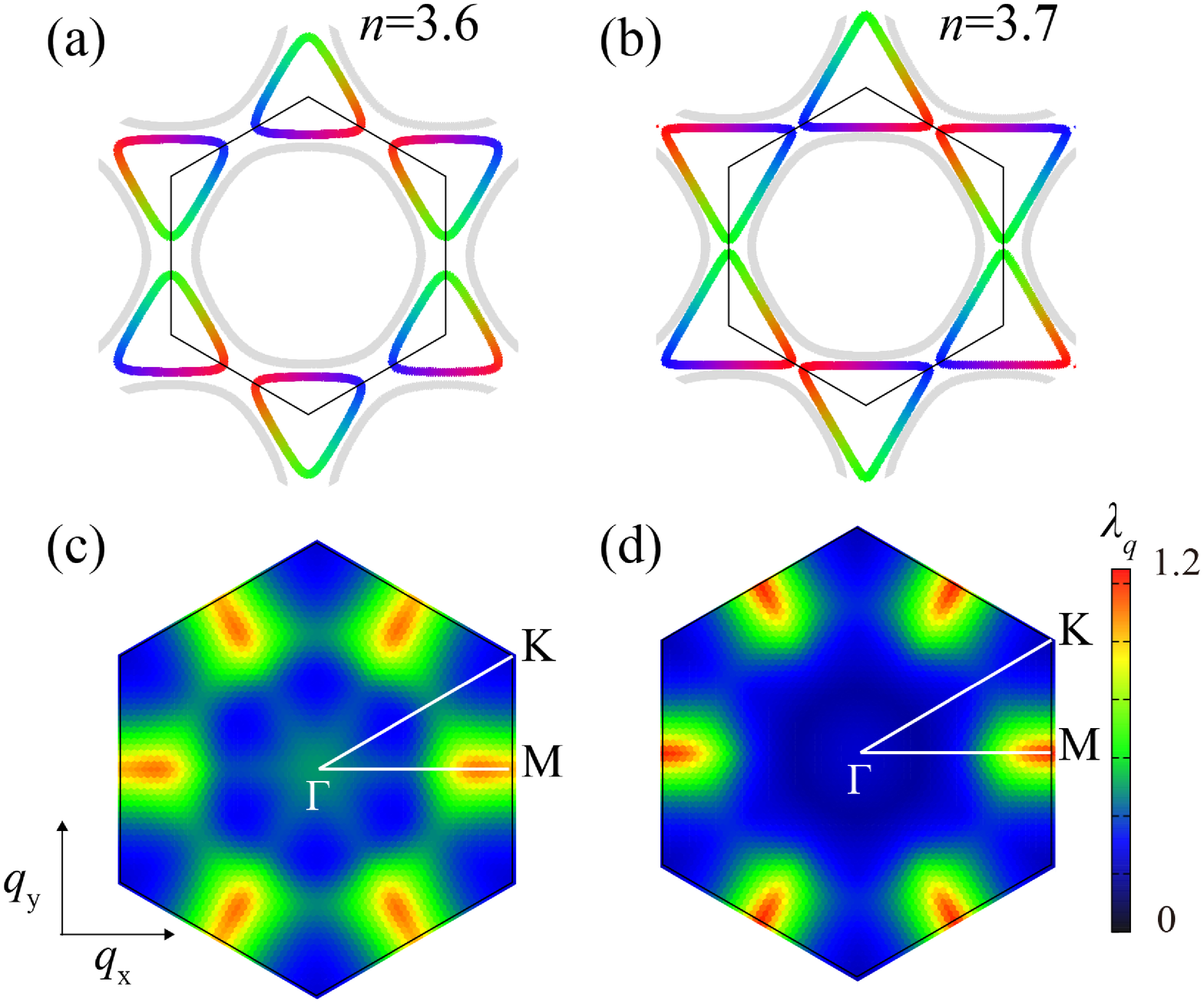}
\caption{
{\bf Robustness of bond-order solution:}
(a) FS for $n=3.6$ and (b) FS for $n=3.7$.
(c) $\q$-dependence of the eigenvalue for $n=3.6$ (incommensurate)
and (d) that for $n=3.7$ (commensurate).
Here, $U=1.34$ ($1.25$) for $n=3.6$ (3.7).
}
\label{fig:figS3}
\end{figure}

First, we study the case of $n<3.8$.
Figures \ref{fig:figS3} (a) and (b) represent the FS 
at $n=3.6$ and $3.7$, respectively.
The obtained eigenvalue of the DW equation at $n=3.6$ and $3.7$
in the case of $\a_S=0.8$ is shown in Figs. \ref{fig:figS3} (c) and (d), 
respectively.
Thus, the commensurate-incommensurate (C-IC) transition occurs 
between $n=3.7$ and $3.6$.
In both cases, the smectic bond-order at $\q_n$ ($n=1,2,3$)
is satisfactorily obtained,
irrespective of the Lifshitz transition at $n_{\rm vHS}=3.71$.
Therefore, we conclude that the strong electron correlation 
due to the three vHS points is essential
for the formation of the bond-order, 
while the shape and the topology of FS are not essential. 
We stress that the second-largest eigenvalue
is much smaller than the present bond-order eigenvalue at $\q=\q_1$.

\begin{figure}[htb]
\includegraphics[width=.9\linewidth]{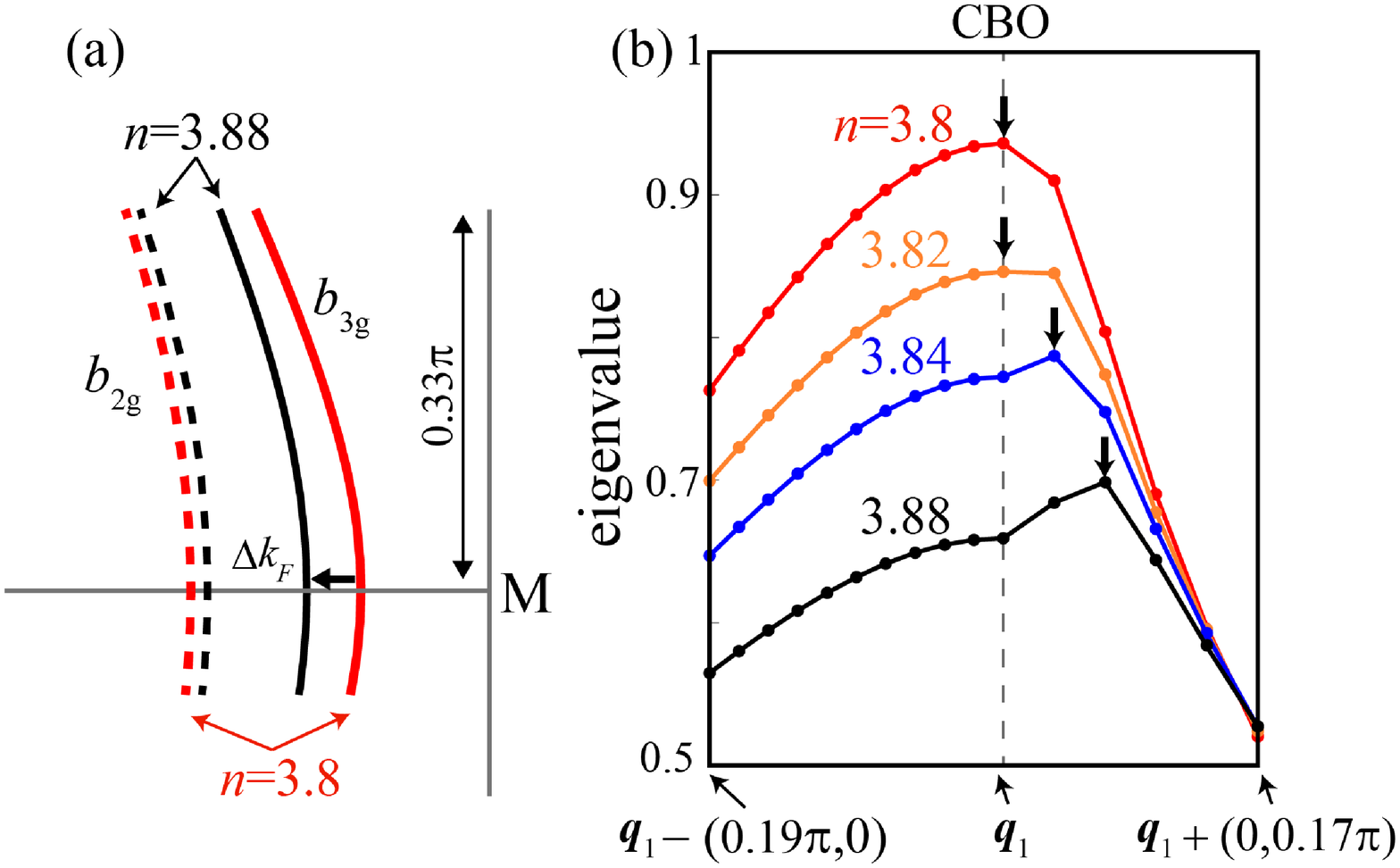}
\caption{
{\bf C-IC bond-order transition in the six orbital model:}
(a) $n$-dependence of the FS in the six orbital model around M point.
(The whole FS at $n=3.8$ is shown in Fig. \ref{fig:fig1} (a).)
The reduction in the $b_{3g}$-orbital
Fermi momentum is expressed as $\Delta k_{\rm F}$.
Here, CBO means the commensurate bond-order.
(b) Obtained $\q$-dependence of the eigenvalue for bond-order
at $T=0.01$ and $\a_S=0.8$.
The C-IC transition occurs at $n=n_0\equiv3.82$.
The wavevector of the DW state is $\q_{DW}=\q_1+(0,\delta)$,
and $\delta>0$ for $n>n_0$.
Here, $U=1.18 \ (1.21)$ at $n=3.80 \ (3.88)$.
}
\label{fig:figS3-2}
\end{figure}

Next, we study the case of $n\ge3.8$.
Figures \ref{fig:figS3-2} (a) and (b) exhibits the 
FSs and the $\q$-dependence of the bond-order eigenvalue $\lambda_\q$,
respectively, in the present simple six orbital model at $T=0.01$.
$U$ is set to satisfy $\a_S=0.80$ at each $n$.
For $n \le n_0\equiv 3.82$, the wavevector of the bond-order is commensurate:
$\q_{\rm DW}=\q_1\equiv (\frac{2}{\sqrt{3}}\pi,0)$.
For $n>n_0$, it changes to incommensurate at $\q_{\rm DW}=\q_1+(0,\delta)$.
The realized electron-doping in the $b_{3g}$-orbital FS is $\Delta n_{b_{3g}}=0.7(n-n_0)$.
The induced shift of the $b_{3g}$-orbital Fermi momentum on the $k_x$-axis
is $\Delta k_{\rm F}=-0.62\pi\times \Delta n_{b_{3g}}$.
Thus, $\Delta k_{\rm F}=-0.02\pi$ is realized when $\Delta n_{b_{3g}}=0.033$ 
(or $n=n_0+0.046$).
On the other hand, in the realistic 30 orbital model in SM G,
$\Delta k_{\rm F}=-0.02\pi$ is induced by the self-doping ($\sim1.5$\%) at $P=3$GPa.
Thus, the C-IC bond-order transition can also be understood
based on the present simple six orbital Hubbard model.

\begin{figure}[htb]
\includegraphics[width=.7\linewidth]{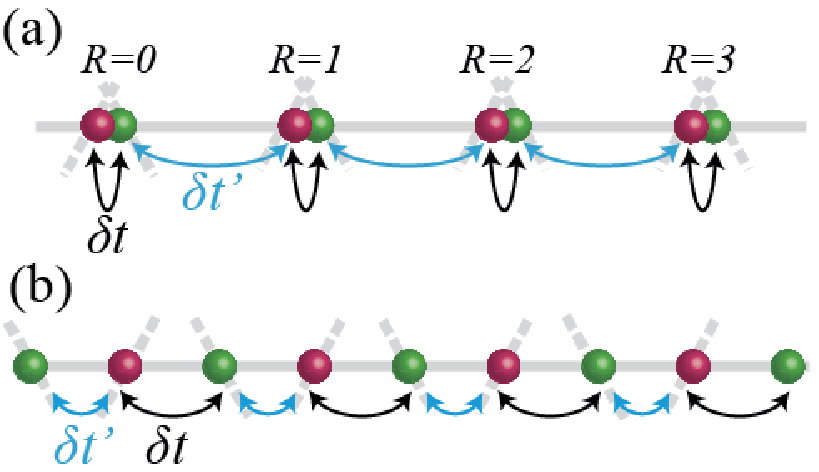}
\caption{
{\bf Origin of bond-order solution in kagome metal:}
(a) Nearest-neighbor hopping modulations
$\delta t$ and $\delta t'$ under the bond-order at $\q=\q_3$
in the triangular lattice model in Fig. \ref{fig:figS1}.
(b) $\delta t$ and $\delta t'$ in the kagome lattice model
in Fig. \ref{fig:fig1} (b). 
}
\label{fig:figS4}
\end{figure}

These numerical results indicates the importance of the vHS points.
Hereafter, we discuss a simplified DW equation
by focusing on the vHS points
in order to understand why bond-order is obtained.
For the bond-order at $\q=\q_3$,
only vHS points A and C are essential,
so we consider a simple two-component form factor
$(f_1,f_2)\equiv (f^{CA}_{\q_3}(\k_{\rm A}),f^{AC}_{\q_3}(\k_{\rm C}))$.
Then, the DW equation at $\q=\q_3$ is given as
\begin{eqnarray}
\lambda 
\begin{pmatrix}
f_1  \\
f_2  \\
\end{pmatrix}
\sim 
N(0)
\begin{pmatrix}
g_{\rm back} & g_{\rm um} \\
g_{\rm um} & g_{\rm back} \\
\end{pmatrix}
\begin{pmatrix}
f_1  \\
f_2  \\
\end{pmatrix},
\label{eqn:vHS-model}
\end{eqnarray}
where $N(0)$ is the density-of-states at the Fermi level.
As we explain in the main text,
both $g_{\rm back}$ and $g_{\rm um}$ are positive.
Thus, the largest eigenvalue and the eigenvector are
$\lambda\sim N(0)(g_{\rm back}+g_{\rm um})$ 
and ${\bm f}=(1,1)$, respectively.
After the Fourier transformation, 
the real-space form factor in the triangular lattice model 
in Fig. \ref{fig:figS1} is given as
$\delta t_{CA}(R{\bm e}_\perp)\sim f_{CA}(\k_{\rm A})e^{i\pi R}$ and 
$\delta t_{AC}(R{\bm e}_\perp)\sim f_{AC}(\k_{\rm C})e^{i\pi R}$.
Here, $R$ is an integer, and 
${\bm e}_\perp$ is a unit vector perpendicular to $\q_3$.
By making comparison between  
Fig. \ref{fig:figS1} and Fig. \ref{fig:fig1} (b) in the main text,
the nearest-neighbor hopping modulations
in Fig. \ref{fig:figS4} (a) are given as $(\delta t, \delta t') =
(\delta t_{CA}({\bm 0}),\delta t_{AC}({\bm e}_\perp))\propto (1,-1)$.

The same bond-order in the kagome lattice model
is shown in Fig. \ref{fig:figS4} (b).
When $\delta t=-\delta t'$,
it is equivalent to Fig. \ref{fig:fig2} (d) in the main text.
Therefore, the essential origin of the bond-order is naturally understood
based on a simple two vHS model in Eq. (\ref{eqn:vHS-model}).

\subsection{C: Unfolded Fermi surface under triple-$\q$ state}

\begin{figure}[htb]
\includegraphics[width=.5\linewidth]{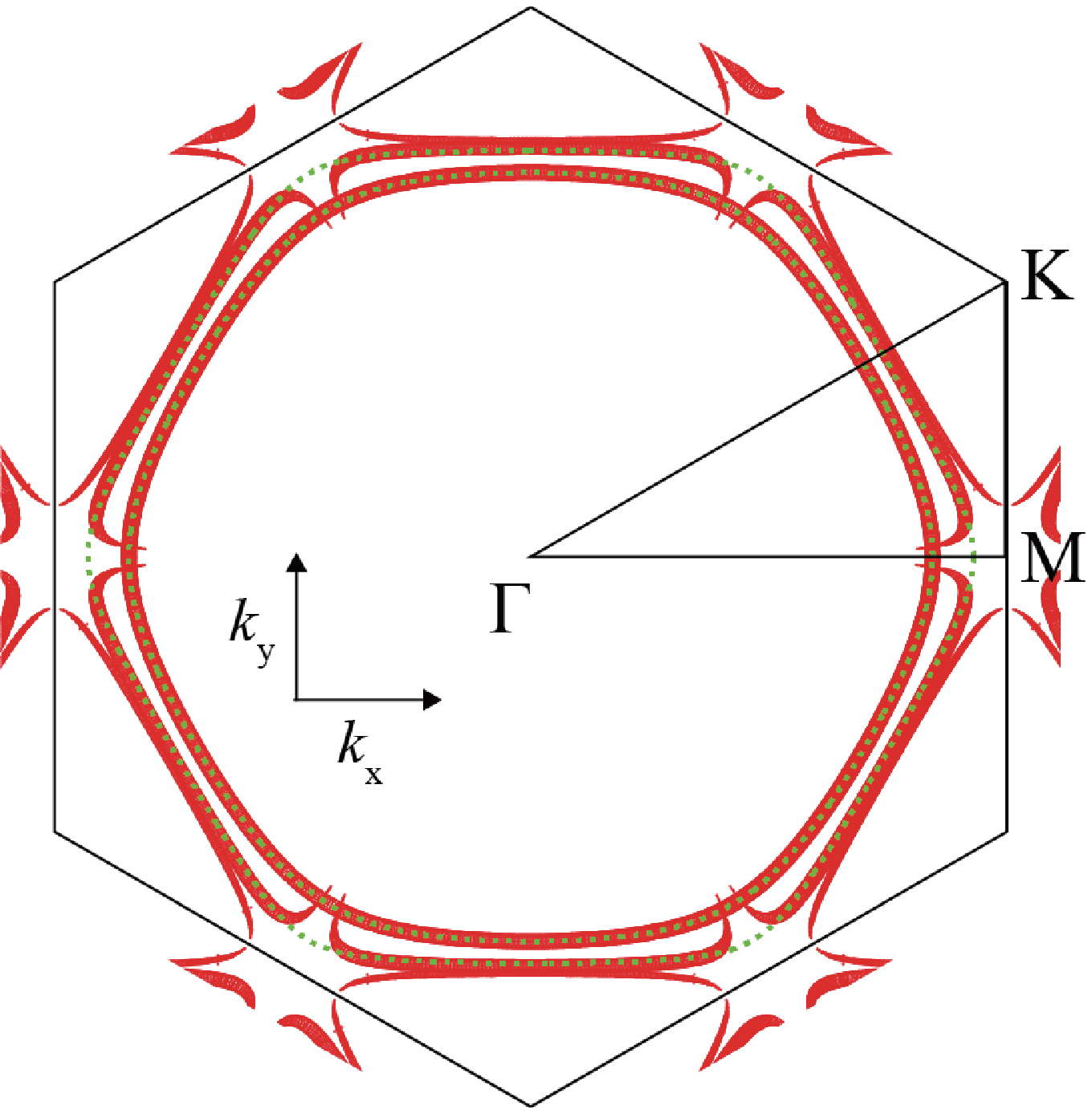}
\caption{
{\bf Unfolded FS under the triple-$\q$ bond-order:}
The FS near the vHS points is reconstructed by the bond-order.
The result for $\max_\k \{f_{\q_n}(\k)\}=0.018$ [eV] is shown.
The green dotted lines represent the original FS.
}
\label{fig:figS5}
\end{figure}

In the triple-$\q$ DW state, both the FS and the bandstructure
are folded into the folded BZ.
They can be unfolded into the original size BZ,
which correspond to the ARPES measurement in the DW state.
The obtained unfolded FS in the case of 
$\max_\k \{f_{\q_n}(\k)\}=0.018$ [eV] is shown in Fig. \ref{fig:figS5}.
Here, the spectra around the vHS points are gapped.
This result is consistent with the recent ARPES studies. 

\subsection{D: GL free energy in $D_{6h}$ kagome model}

Here, we briefly review the GL free energy of a $D_{6h}$ system 
up to the fourth order and explain that the triple-$\q$ order is 
stabilized by the third-order term.
We introduce three real order parameters $\phi_n$ ($n=1-3$)
and express the bond-order functions as
$\phi_n {\hat f}_{\q_n}(k)$, 
where ${\hat f}_{\q_n}(k)$ is the normalized dimensionless 
form factor given by the linearized DW equation.
[Note that the phase factor of the form factor is fixed by the
Hermite condition $f_{q}^{lm}(k) = [f_{-q}^{im}(k+q)]^*$.]
The GL free energy is given as \cite{Hirata}
\begin{eqnarray}
F &=& a [\phi_1^2+\phi_2^2+\phi_3^2] + b\phi_1 \phi_2 \phi_3
\nonumber \\
& &+c [\phi_1^4+\phi_2^4+\phi_3^4]
+d [\phi_1^2\phi_2^2+\phi_2^2\phi_3^2+\phi_3^2\phi_1^2]
\end{eqnarray}
where the second-order coefficient $a$ is proportional to 
$1-\lambda_{\rm bond}$.
The fourth-order coefficients $c,d$ are positive.
Here, the third-order coefficient $b$ is nonzero
because of the momentum conservation relation $\q_1+\q_2+\q_3={\bm 0}$.
Note that the sign of $b$ is reversed under the transformation $f \rightarrow -f$.

One can calculate the coefficients $b$, $c$, and $d$ microscopically 
based on their diagrammatic expressions given in our previous paper 
on 1T-TaS$_2$ \cite{Hirata}:
The coefficient $b$ is given by the triangle diagram composed of 
three Green functions and three form factors. 
In a similar way, the coefficients $c$ and $d$ 
are given by the square diagrams.

Based on the GL free energy, 
the single-$\q$ solution $(\phi_1,\phi_2,\phi_3)=(\phi,0,0)$
occurs when $a\le 0$ as the second-order transition.
On the other hand, in the case of the triple-$\q$ order, 
the free energy is $F(\phi)=a\phi^2+b\phi^3/3\sqrt{3}+(c+d)\phi^4/3$.
In the case $a>0$, its local minimum is given at
$\displaystyle \phi_0=\frac{-\sqrt{3}b}{8(c+d)}\left(1-\sqrt{1-\frac{32a}{b^2}(c+d)}\right)$,
which is positive [negative] for $b<0$ [$b>0$].
The free energy $F(\phi_0)$ becomes negative when 
$a<b^2/[36(c+d)]$ (=positive).
Therefore, triple-$\q$ order is realized as the first order transition 
when $a$ is positive.
(That is, $\phi$ jumps from zero to $\phi_0$ at finite $a>0$.) 
Consistently with this analysis, 
the star of David bond-order emerges
as a weak first order transition experimentally.

\subsection{E: Derivation of SC gap equation}

Here, we discuss the reason
why bond-order fluctuations mediate the pairing interaction.
In Ref. \cite{Kontani-PRB2011}, the authors studied
the orbital fluctuation mediated $s$-wave superconductivity 
in Fe-based superconductors.
In that study, the electron-boson coupling (=form factor) 
is an orbital-dependent but $\k$-independent charge quadrupole operator:
$\hat{f}^{\q}(\k)=\hat{O}_\Gamma$ $(\Gamma=xz,yz,xy)$.
In the main text, 
we obtain the development of bond-order fluctuations with the 
$\k$-dependent form factor in AV$_3$Sb$_5$, which is given by the
inter-sublattice vertex corrections (VCs) that are dropped in the RPA.
We reveal that bond-order fluctuations 
mediate significant ``beyond-Migdal'' pairing interaction
thanks to the $\k$-dependent  form factor \cite{Onari-AFN},
and therefore $s$-wave and $p$-wave SC states emerge in AV$_3$Sb$_5$.

\begin{figure}[htb]
\includegraphics[width=.99\linewidth]{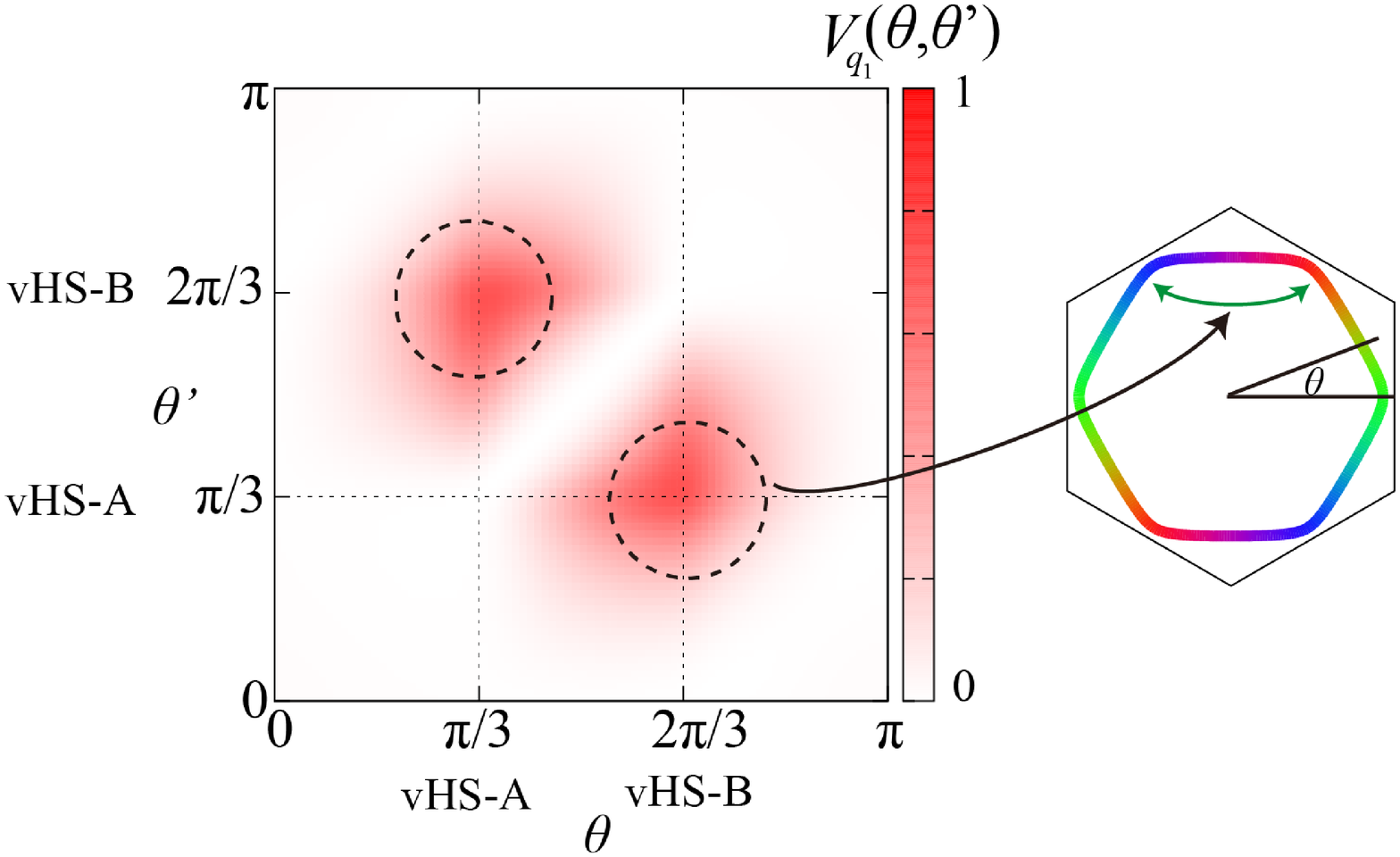}
\caption{
{\bf Pairing interaction on $b_{3g}$-orbital FS due to bond-order fluctuations:}
We present $V_1(\theta,\theta')$ due to $\q\approx\q_1$
given in (\ref{eqn:V_n})
in the case of ${\bar I}_\q=1$ and $\lambda_{\rm bond}=0$.
Here, $\theta=\arctan(k_y/k_x)$ and $\theta'=\arctan(k_y'/k_x')$.
}
\label{fig:figS6}
\end{figure}

In the following, we discuss the pairing interaction due to
the bond-order fluctuations in kagome metal by following Ref. \cite{Onari-AFN}.
Hereafter, we drop the orbital indices just to simplify the notation.
The pairing interaction between Cooper pairs $(k,-k)$ and $(k',-k')$
due to charge-channel full four-point vertex in Fig. \ref{fig:figS2} (c)
is given as $V(k,k')\propto \Gamma_{k'-k}^c(k,-k')$.
We derive a convenient simple expression of $\Gamma_{\q}^c(k,k')$,
we introduce the following approximation for the kernel function:
\begin{eqnarray}
I_\q(\k.\k')= {\bar I}_\q f_\q(k)f_\q^*(k'),
\label{eqn:Iapprox}
\end{eqnarray}
where
$f_\q(k)$ is the form factor for the largest eigenvalue
of the DW equation.
By inserting Eq. (\ref{eqn:Iapprox}) into the DW equation,
the eigenvalue is given as $\lambda_\q= {\bar I}_\q\chi_f^0(\q)$,
where 
\begin{eqnarray}
\chi_f^0(\q) = -\frac{T}{N} \sum_k G(k+\q)G(k) {\tilde f}_\q(k){\tilde f}_\q^*(k).
\ \ (>0) 
\end{eqnarray}
Then, the full four-point vertex is given as
\begin{eqnarray}
\Gamma_{\q}^c(k,k')=\frac{{\bar I}_\q f_\q(k)f_\q^*(k')}{1-\lambda_\q} .
\end{eqnarray}
Therefore, the pairing interaction is 
$V(k,k')= \frac{{\bar I}_q f_\q(k)f_\q^*(-k')}{1-\lambda_\q}$,
where $\q=\k'-\k$.
Note that the relation $f_\q^*(-k')=f_{-\q}(-k)$ holds for $\q=\k'-\k$
due to the Hermite condition of the form factor.

Considering that the $\q$-dependence of the form factor is moderate,
the total pairing interaction due to 
triple-$\q$ bond-order fluctuations is approximately given as
\begin{eqnarray}
V_{\rm bond}(k,k')&=& \frac12 \sum_{n}^{1,2,3} V_n(k,k') ,
\end{eqnarray}
\begin{eqnarray}
V_n(k,k')&=& \frac{{\bar I}_\q 
{\bar f}_{\q_n}(\k){\bar f}_{\q_n}^*(-\k')}{1-\lambda_{\k'-\k}}
 \nonumber \\ 
&\approx& \frac{{\bar I}_{\rm bond} 
{\bar f}_{\q_n}(\k){\bar f}_{-\q_n}(-\k)}{1-\lambda_{\rm bond}}
\frac1{1+\xi^2(\q_n-(\k'-\k))^2} ,
\nonumber \\
\label{eqn:V_n}
\end{eqnarray}
where
${\bar f}^\q(\k)\equiv \sum_{l,m}{\tilde f}^\q_{lm}(\k) u_{l,b}(\k+\q)^* u_{m,b}(\k)$.
Here, $u_{l,b}(\k)=\langle l,\k| b,\k\rangle$ is the unitary transformation matrix element 
between orbital $l$ and conduction band $b$,
and ${\bar f}^\q(\k)\equiv({f}^\q(\k,\pi T)+{f}^\q(\k,-\pi T))/2$.
We also approximate $\lambda_\q \approx \lambda_{\rm bond} -b(\q-\q_n)^2$
with $b\approx \xi^2 (1-\lambda_{\rm bond})$ for $\q\sim\q_n$.
Here, we set ${\bar f}_{\q_1}^{BA}(k_{\rm B})=1$.
Then, the coupling constant ${\bar I}_{\rm bond}$ is 
directly given by $g_{\rm um}$ that is obtained in the main text.
We stress that ${\bar f}_{\q_n}(k){\bar f}_{-\q_n}(-k)$
is positive for even-parity bond-order.
In the main text, we set $\bar{I}_{\rm bond}=g_{\rm um}$ 
$|\bar{f}_{\q_1}(\k_{\rm A})|=1$ 
in the pairing interaction in Eq. (\ref{eqn:V_n}).

Figure \ref{fig:figS6}
is the pairing interaction $V_1(\theta,\theta')$ on the $b_{3g}$-orbital FS
due to the bond-order fluctuations at wavevector $\q\approx\q_1$.
by setting ${\bar I}_\q=1$, $\lambda_{\rm bond}=0$ and $\xi=0$.
Here, $\theta=\arctan(k_y/k_x)$ and $\theta'=\arctan(k_y'/k_x')$.
Thus, strong attractive pairing interaction is induced by the bond-order fluctuations 
around the vHS points.
This is the driving force of the $s$-wave and $p$-wave pairing states
obtained in the main text.

In the main text, we solve the gap equation in the presence of 
bond and spin fluctuations.
\begin{eqnarray}
\!\!\!\!\!\!\!\!\!\!
V_{\rm s}^{\rm SC}(k,k')&=& V_{\rm bond}(k,k') -\frac{3}{2} U^2\chi^s(k-k')-U ,
\label{eqn:S-Vs} \\
\!\!\!\!\!\!\!\!\!\!
V_{\rm t}^{\rm SC}(k,k')&=& V_{\rm bond}(k,k') +\frac{1}{2} U^2\chi^s(k-k') ,
\label{eqn:S-Vt} 
\end{eqnarray}
where s (t) represents the singlet (triplet) pairing interaction.
The diagrammatic expression of the gap equation due to 
$V_{\rm bond}(k,k')$ is depicted in Fig. \ref{fig:fig4} (a)
in the main text.
In solving the gap equation, 
we set the BCS cutoff energy $\w_c$ for $V_{\rm bond}$
because the energy-scale of bond-order fluctuations 
is much smaller than $E_F$.
Here, we set $\w_c=0.02$.
Note that the pairing interaction for the band $b,b'$
can be derived from that in the orbital representation 
by using the unitary transformation matrix $u_{l,b}(\k)$.


It is noteworthy that 
the present bond-order fluctuating pairing mechanism 
is outside of the Migdal approximation, in which the 
form factor is assumed to be $\k$-independent.
The present bond-order fluctuating mechanism has a close similarity 
to the multiple-fluctuation pairing mechanism 
developed in Refs. \cite{Onari-AFN,Tazai-HF-SC1}.

\subsection{F: Comparison between the present DW equation theory and mean-field theory} 

\begin{figure}[htb]
\includegraphics[width=.99\linewidth]{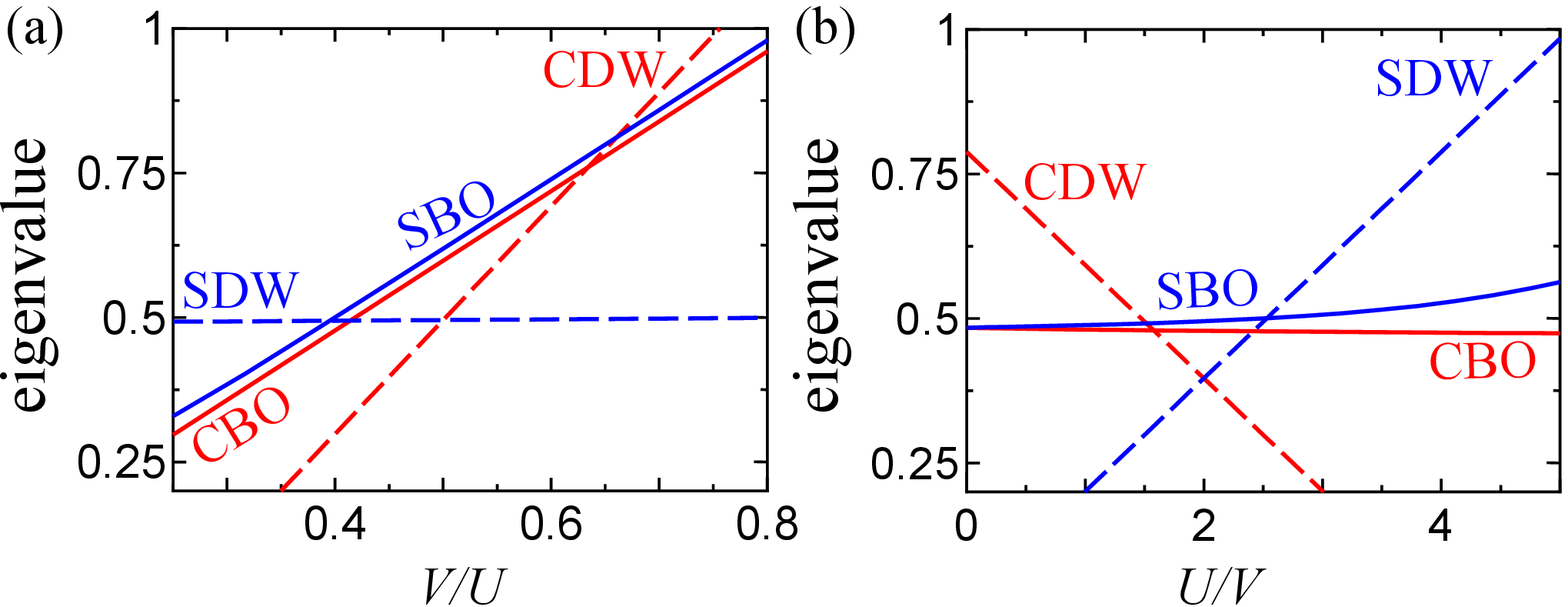}
\caption{
{\bf Mean-field analysis for $U$-$V$ Hubbard model}:
(a) Eigenvalues of the mean-field density-wave equation 
as a function of $V/U$ at $U=0.79$, and 
(b) those as a function of $U/V$ at $V=0.32$.
Here, the charge-BO instability is smaller than 
other instabilities.
}
\label{fig:figS7}
\end{figure}

A great merit of the present paramagnon-interference theory is that 
the star of David bond-order is naturally obtained based on a simple Hubbard model 
without introducing any off-site Coulomb interactions.
Here, we briefly review the results of the mean-field theory based on the $U$-$V$ Hubbard model,
where $V$ is the nearest-neighbor Coulomb interaction.
For this purpose, 
we solve the linearized mean-field equation with the 
optimized form factor, which is given by 
Eq. (2) with the Hartree-Fock kernel function made of $U$ and $V$.
Figure \ref{fig:figS7} (a) shows the 
obtained eigenvalues as a function of $V/U$ at $U=0.79$.
(Here, we drop $b_{2g}$ orbitals in the kagome metal Hubbard model introduced in the main text,
because $b_{2g}$ orbitals are not essential for the bond-order.)
The spin-density-wave (SDW) instability 
(with spin form factor $f^s=1$) is the largest for $V\lesssim0.4U$.
The SDW is replaced with the spin-bond-order (spin-BO), $f^s\ne 1$) for
 $0.4U \lesssim V \lesssim 0.6 U$, 
and the charge-density-wave (CDW) instability 
(with charge form factor $f^c=1$) is the largest for $V\gtrsim0.6U$. 
Figure \ref{fig:figS7} (b) shows the 
results as functions of $U/V$ at $V=0.32$.
Thus, the charge bond-order (charge-BO) instability is secondary 
in both Figs. \ref{fig:figS7} (a) and (b).

\begin{figure}[htb]
\includegraphics[width=.7\linewidth]{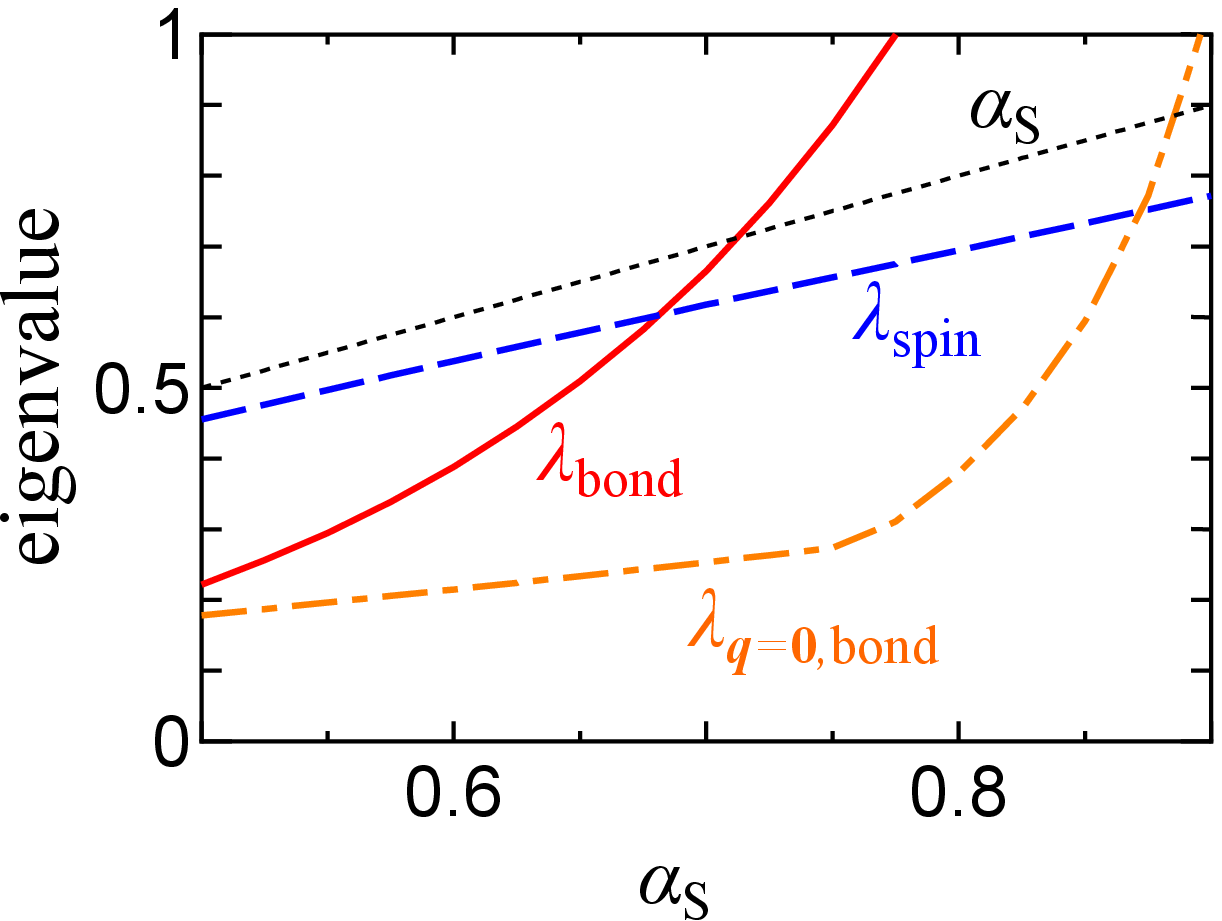}
\caption{
{\bf DW equation analysis for $U$ Hubbard model}:
Eigenvalue of DW equation with MT and AL terms as a function of $\a_S \ (\propto U)$.
Here, the eigenvalue of charge-channel bond-order 
$\lambda_{\rm bond}\equiv \lambda_{\q_1}^c$ solely increases due to the AL terms.
In contrast, the spin-channel eigenvalue $\lambda_{\rm spin}\equiv \lambda_{\q_1}^s$
remains small even if the spin-channel vertex corrections are taken into account.
Interestingly, $\lambda_{\q={\bm0},{\rm bond}}\equiv \lambda_{\q={\bm0}}^c$,
which corresponds to $A_{1g}$ bond-order with sign-reversing form factor,
starts to develop for $\a_S\gtrsim 0.8$.
}
\label{fig:figS8}
\end{figure}

In contrast, the charge bond-order instability is solely magnified 
in the DW equation with MT and AL terms even for $V=0$.
Figure \ref{fig:figS8} exhibits the spin-channel and charge-channel 
eigenvalues, $\lambda_\q^{s}$ and $\lambda_\q^{c}$,
in the DW equation at $\q={\bm0}$ and $\q=\q_1$
\cite{Kontani-sLC}.
We see that $\lambda_{\q=\q_1}^{c}$,
which is equivalent to $\lambda_{\rm bond}$ in the main text, 
solely increases with increasing $\a_S \ (\propto U)$.
Thus, the bond-order solution is obtained for wide parameter range.
Interestingly, $\lambda_{\q={\bm0},{\rm bond}}\equiv \lambda_{\q={\bm0}}^c$,
which corresponds to $A_{1g}$ bond-order with sign-reversing form factor,
starts to develop for $\a_S\gtrsim 0.8$.

By solving the spin-channel DW equation \cite{Kontani-sLC},
we reveal that the spin channel eigenvalue 
$\lambda_{\rm spin}\equiv \max_\q\lambda_{\q}^{s}$
is smaller than $\a_S$ only slightly,
and it corresponds to the SDW solution with $f^s\approx 1$.
(Note that $\lambda_{\rm spin}=\a_S$ in the mean-field approximation for $V=0$.)
We stress that the spin-BO ($f^s\ne 1$) instability is always smaller 
than the SDW instability.
In the present theory, the charge-channel AL term 
is proportional to the convolution
$C^c(\q_1) = \sum_\k \chi^s(\k+\q_1)\chi^s(\k)$,
while the spin-channel one is proportional to
$C^s(\q_1) = \sum_\k \chi^s(\k+\q_1)\chi^c(\k)$,
as we discussed in Refs. \cite{Yamakawa-FeSe,Kontani-sLC}.
Since $C^c(\q_1) \gg |C^s(\q_1)|$ for $\a_S\gtrsim 0.75$,
only the charge bond-order eigenvalue $\lambda_{\rm bond}$ 
is strongly enlarged and exceeds unity in Fig. \ref{fig:figS8}.
Thus, the instability of the smectic bond-order is robust in kagome metals,
while any spin-channel eigenvalues are smaller than unity
even if the quantum interference mechanism is taken into account.
This is a great merit of the present DW equation analysis.

\subsection{G: Realistic model Hamiltonian based on the first-principles study} 

\begin{figure}[htb]
\includegraphics[width=.87\linewidth]{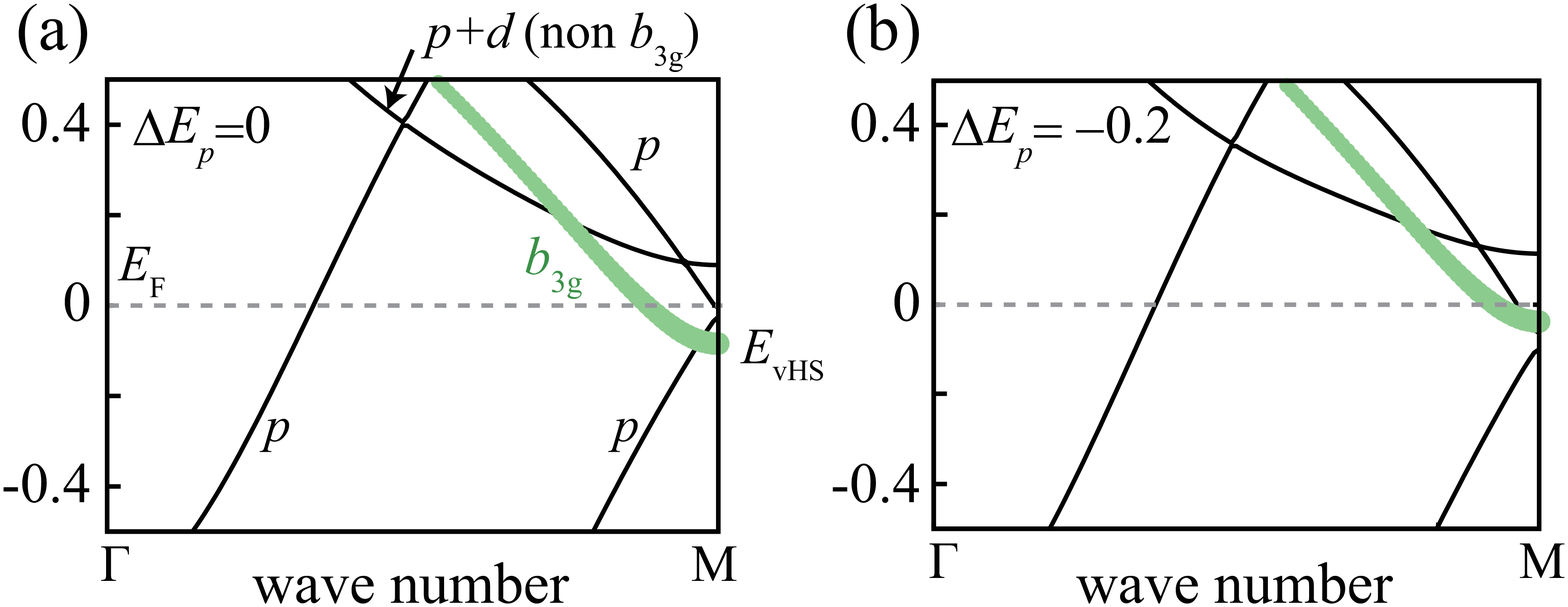}
\caption{{\bf Band structure based on the first-principles calculation}:
(a) Original DFT band structure for CsV$_3$Sb$_5$ in the case of $\Delta E_p=0$.
(b) Modified band structure with $\Delta E_p=-0.2$ and 0.1 hole-doping. 
}
\label{fig:band_PT}
\end{figure}

Based on the first-principles study,
we derive the two-dimensional realistic model for CsV$_3$Sb$_5$,
which we analyzed in the main text.
First, we perform the Wien2k DFT calculation of CsV$_3$Sb$_5$ under pressure $P$ [GPa], 
using the crystal structure data in Ref. \cite{1stprin}.
Next, we derive the 30 orbital (15 3$d$-orbitals + 15 5$p$-orbitals) 
tight-binding model, $H_0^{\rm DFT}(P)$, using the Wannier90 software.
The DFT band dispersions are fitted almost perfectly.
Here, we drop the inter-layer hopping integrals.
The bandstructure along the $k_x$-axis 
at $P=0$ is shown in Fig. \ref{fig:band_PT} (a).
The green line is the $b_{3g}$-orbital band, and its width represents the 
$b_{3g}$-orbital weight.
The bandstructure in Fig. \ref{fig:band_PT} (a) is 
qualitatively similar to ARPES data.
However, Fig. \ref{fig:band_PT} (a) is 
different from the experimental CsV$_3$Sb$_5$ bandstructure
\cite{STM1,ARPES-VHS,ARPES-band}
at the following two points:
(i) The vHS energy $E_{\rm vHS}$ is not adjacent to 
the Fermi level $E_{\rm F}$.
(ii) The $b_{3g}$-band crosses a 5$p$-band around M point,
so the $b_{3g}$-orbital weight disappears 
via the band hybridization except on the $k_x$-axis.

We will fix these discrepancies
by introducing {\it minimum changes} in the Hamiltonian.
Here, we shift 15 $5p$-orbital levels by $\Delta E_p=-0.2$ [eV]
and introduce the hole-doping by 0.1 to make $E_{\rm vHS}$ closer to $E_{\rm F}$.
(The orbital-dependent energy shift method is frequently applied
in Fe-based superconductors to reproduce experimental FSs 
\cite{Yamakawa-FeSe}.)
The obtained modified bandstructure is shown in Fig. \ref{fig:band_PT} (b)
and its FS is presented in Fig. \ref{fig:figPT} (a),
which are consistent with recent
STM and ARPES measurements \cite{STM1,ARPES-VHS,ARPES-band}.
We use this model as the model Hamiltonian at ambient pressure, $H_0(P=0)$.

Next, we discuss the pressure dependence of the Hamiltonian
caused by the systematic change in the crystal structure.
It is described by the 
``pressure Hamiltonian $\Delta H_0^{\rm DFT}(P)\equiv H_0^{\rm DFT}(P)-H_0^{\rm DFT}(0)$''.
Then, the model Hamiltonian is given as 
\begin{eqnarray}
H_0(P)= H_0(0)+\Delta H_0^{\rm DFT}(P).
\label{eqn:H0P}
\end{eqnarray}
This method has been successfully applied to 
the study of $P$-$T$ phase diagram in Fe-based superconductors.
The obtained FSs at $P=0$ and 3GPa are shown in 
Figs. \ref{fig:figPT} (a) and (b).
Under pressure, the volume of $b_{3g}$-orbital FS is reduced 
because the $b_{3g}$-orbital level shifts downward relatively.
The self-doping on the $b_{3g}$-FS ($\sim 1.5$\% at $P=3$GPa)
derived from $\Delta H_0^{\rm DFT}(P)$ will be reliable.
At the same time, the bandwidth increases under pressure, and 
the spin Stoner factor $\a_S$ is reduced by $0.03$ at $P=3$GPa.
(In this model, $\a_S=0.95$ at $U=2.7$ [eV] when $P=0$.)
By using Eq. (\ref{eqn:H0P}),
the pressure-induced C-IC bond-order transition 
is explained in the main text.
It is an important future problem 
to analyze three-dimensional Hubbard model of CsV$_3$Sb$_5$.




\end{document}